\documentclass[superscriptaddress,showkeys,twocolumn,nofootinbib,longbibliography]{revtex4-1}

\usepackage{amsmath}
\usepackage{amssymb}
\usepackage{graphicx}
\usepackage{epsf}
\usepackage{slashed}
\usepackage{enumitem}
\usepackage[czech,english]{babel}
\usepackage[cp1250]{inputenc}
\usepackage{hyperref}
\usepackage{xcolor}

\usepackage[only,llbracket,rrbracket,llparenthesis,rrparenthesis]{stmaryrd} % for comparison with 4 similar parenthesis symbols
\usepackage{accsupp} % for ensuring the right Unicode codepoint upon pasting

\newcommand{\refer}[1]{(\ref{#1})}
\newcommand{\diff}{\mathrm{d}}

\newcommand{\lineint}{\int\limits_{-\infty}^{\infty}\diff}

\newcommand{\Exp}[1]{\mathrm{e}^{#1}}

\newcommand{\iunit}{\mathrm{i}}

\usepackage{bbm} %\mathbbm{1}, \mathbbmss{1}, \mathbbmtt{1}

\usepackage[nodayofweek]{datetime}
\newdateformat{mydate}{\twodigit{\THEDAY}{ }\shortmonthname[\THEMONTH], \THEYEAR}
\usepackage[normalem]{ulem}
\begin{document}

\title{Mechanization of scalar field theory in 1+1 dimensions}

\author{Filip Blaschke}
\email{filip.blaschke(at)fpf.slu.cz}
\affiliation{
Research Centre for Theoretical Physics and Astrophysics, Institute of Physics, Silesian University in Opava, Bezru\v{c}ovo n\'am. 1150/13, 746~01 Opava, Czech Republic. \\
and Institute of Experimental and Applied Physics, Czech Technical University in Prague, Husova 240/5, 110 00 Prague 1, Czech Republic
}

\author{Ond\v{r}ej Nicolas Karp\'{i}\v{s}ek}
\email{karponius(at)gmail.com}
\affiliation{
Faculty of Philosophy and Science in Opava, Silesian University in Opava, Bezru\v{c}ovo n\'am. 1150/13, 746~01 Opava, Czech Republic,
}

\begin{abstract}
The `mechanization' is a procedure of replacing a scalar field in 1+1 dimensions with a piece-wise linear function, i.e. a finite graph consisting of $N$ joints (vertices) and straight segments (edges). As a result, the field theory is approximated by a sequence of algebraically tractable, general-purpose collective coordinate mechanical models. 
We observe the step-by-step emergence of dynamical objects and associated phenomena as the $N$ increases. Mech-kinks and mech-oscillons -- mechanical analogs of kinks and oscillons (bions) -- appear in the simplest models, while more intricate dynamical patterns, such as bouncing phenomenon and bion pair-production, emerge gradually as decay states of high $N$ mech-oscillons.
%\remark{Version V1, compiled: \mydate\today, \currenttime.}
\end{abstract}


\maketitle

%\tableofcontents

%\raggedbottom

\section{Introduction}
\label{sec:I}

%long live the kink
Scattering of solitons provides a unique window to the inner workings of non-linear field theories. Even in the case of the simplest solitons -- kinks -- the rich phenomenology that we see has not been completely understood despite four decades of theoretical research \cite{Campbell:1983xu, Moshir:1981ja,  Belova:1985fg, Anninos:1991un, Halavanau:2012dv, Dorey:2011yw, Weigel:2013kwa, Haberichter:2015xga, Ashcroft:2016tgj, Manton:2020onl, Manton:2021ipk,  Adam:2019prh, Adam:2021gat, 2019arXiv190903128K}.

The prototypical non-linear field theory is the $\phi^4$ model with a single scalar field and double-well potential $V(\phi) = \tfrac{1}{2}\bigl(1-\phi^2\bigr)^2$, where the kink solution has an especially simple form: $\phi_K(x) = \tanh(x)$.
Numerical investigations of kink ($K$) anti-kink ($\bar K$) collisions reveal an intricate pattern of possible outcomes \cite{Campbell:1983xu, Moshir:1981ja, Belova:1985fg, Anninos:1991un, Adam:2021gat, Manton:2021ipk}: i) Above the critical velocity $v_{\rm crit} \approx 0.26$, the colliding pair is immediately reflected and escapes to infinity. ii) Below this threshold, the solitons sometimes form a `bound state' (bion) that slowly decays to the vacuum via the emission of radiation. iii) Before final reflection and escape to infinity,  the kink and anti-kink bounce off each other several times. These bounces occur in narrow windows of initial velocities that are nested and fractal-like  punctuated by `bion chimneys' and are observed only for initial velocities larger than $v_{\rm min} \approx 0.18$. 

To understand these phenomena is, at its core, a problem of complexity. During the $K\bar K$ collisions, 
the field enters a deeply non-linear regime, where our analytic and `weak-field' perturbative tools fail. It could be the case that a very large number of degrees of freedom participate with equal importance, and we may never untangle the web of inner-dependencies. That would be a scenario of \emph{computational irreducibility} that is characteristic for some discrete dynamical systems, such as cellular automata \cite{Wolf}. Fortunately, for solitons it has been an overwhelming experience that the opposite is true -- there is a surprisingly large \emph{reducibility}. Namely, a deep understanding of the dynamics seems to be possible through tracking only a few effective degrees of freedom, called \emph{collective 
coordinates} (CCs).

This is most evident for multi-soliton configurations that are BPS \cite{Bogomolny:1975de, Manton:2004tk}. The small-velocity scattering of BPS solitons can be accurately described using the moduli (geodetic) approximation, where the CCs are time-dependent parameters (moduli) of the static solution, such as mutual separations, etc. Integrating the field's kinetic energy for this background we get a metric of the moduli space -- a curved manifold of BPS solutions -- and  the scattering of solitons is equivalent to the geodetic motion \cite{Manton:1981mp}.  

For a non-BPS case, the canonical moduli space does not exist and we are forced to guess.
However, a general strategy, which can be perhaps called an \emph{adiabatic} method, uses the same principle, namely it introduces an ad hoc moduli space that aspires to capture the dynamics under investigation. The coordinates of this space become dynamical variables in the resulting Collective Coordinate Model (CCM), usually resembling classical mechanics of interacting point masses.

In the case of $\phi^4$ kinks, there are no static multi-kink solutions (however, one can stabilize them by adding impurities \cite{Adam:2019prh}). The use of CCM for  $\phi^4$ kinks has a long, fruitful, and somewhat complicated history (see \cite{2019arXiv190903128K} for a recent overview).
For our purposes, let us highlight two important results. 

First, for a single kink, it has been noticed a long-time ago \cite{RiceMJ} that a correct CCM (in the sense of both qualitative and quantitative agreement)  must include not only the position modulus but also a scaling modulus. If we plug the ansatz
\begin{equation}
\phi_{\rm bkg} = \tanh\bigl(b(t)(x-a(t))\bigr)
\end{equation}
into the Lagrangian and integrate over $x$, the resulting effective theory (Eq.~\refer{eq:bpskinklag}), while non-linear, can be exactly integrated and contain both Lorentz-boosted kink solution and exact harmonic motion, intimately connected with the so-called \emph{Derrick mode} of the kink \cite{Manton:2020onl}. This mode has been recognized as a key tool for recovering Lorentz covariance that is typically lost in CCMs.

The second result is the recent identification of quantitatively correct CCM for $K\bar K$ scattering \cite{Manton:2020onl, Manton:2021ipk, Adam:2021gat} that is based on the observation that a simple $K\bar K$ superposition ansatz
\begin{equation}\label{eq:twokinkans}
\phi_{\rm bkg} = \tanh\bigl(b(t)(x+a(t))\bigr)-\tanh\bigl(b(t)(x-a(t))\bigr)-1\,,
\end{equation}
leads to moduli space with unremovable singularity at the point $a=0$. Therefore, in \cite{Adam:2021gat} the authors proposed a perturbative approach for incorporating the scaling modulus order by order. The resulting CCMs do not suffer from the singularity at $a=0$ and seem to qualitatively and quantitatively match observed $K\bar K$ dynamics. 

The use of CCMs is, of course, not limited just to $K\bar K$ scattering, but it can be also applied also on other problems, such as the lifetime of an oscillon. 
A field-theoretical oscillon is a meta-stable state that is observed in the collapse of bubbles -- usually taken as Gaussian peaks placed on top of a vacuum. Compared with models with only a single vacuum, when the scalar field theory has two or more vacua, the lifetime of such bubbles is typically much longer than one would naively expect. Further, certain universal features are shared across various initial conditions and models. At the onset, the initial profile quickly settles to a certain shape around which it oscillates quasi-periodically for a long time slowly radiating away energy, until it suddenly decays. The oscillons enjoy a long history of investigation, although typically in spatial dimensions higher than one \cite{Gleiser:1993pt, Copeland:1995fq, Andersen:2012wg, Salmi:2012ta, Saffin:2006yk, Fodor:2006zs, Gleiser:2009ys, Olle:2020qqy}.

\begin{figure}[htb!]
\begin{center}
\includegraphics[width=0.95\columnwidth]{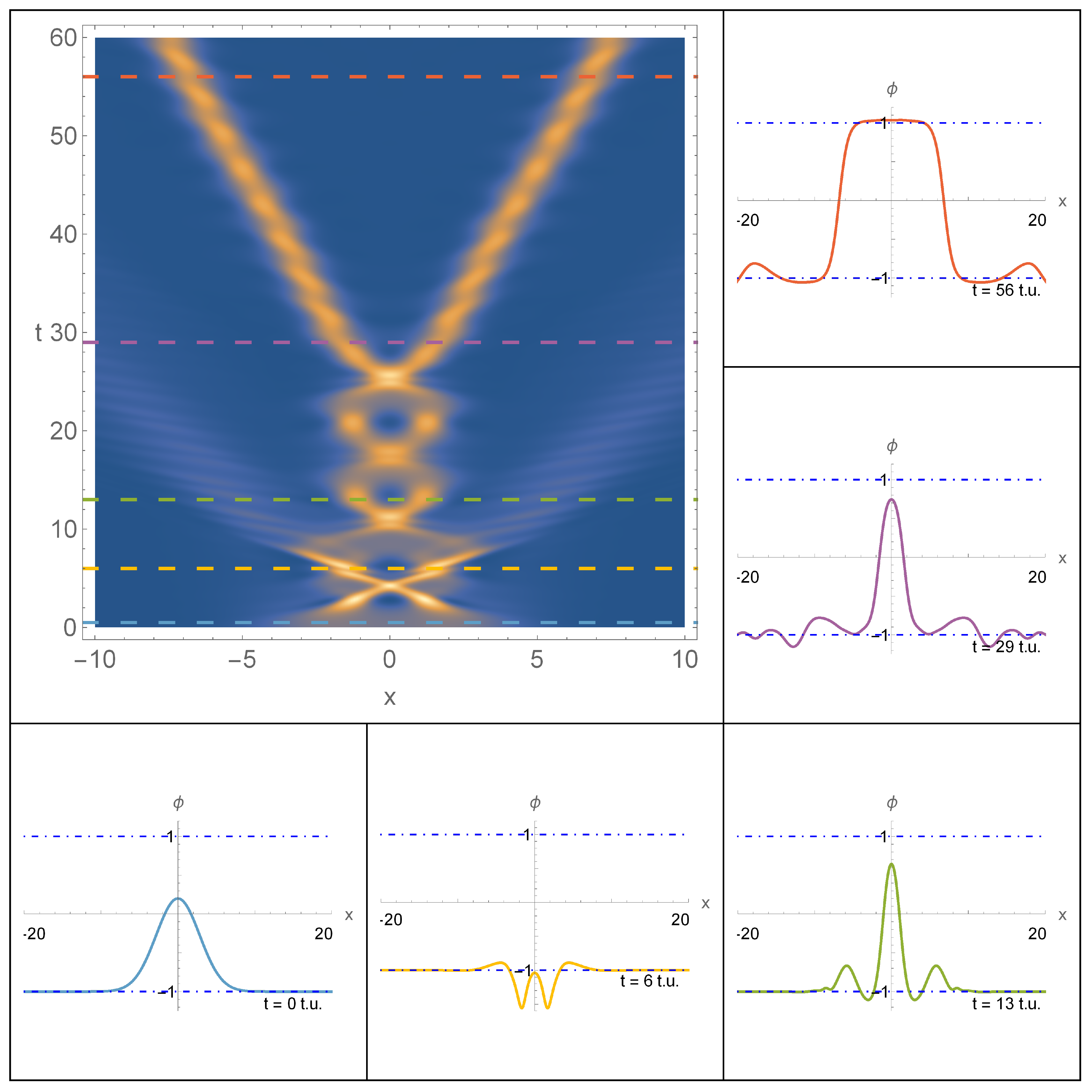}
\end{center}
{
    \caption{\small The plot of the energy density and field profiles for several timestamps in the $\phi^4$ model showcasing a disintegration of a Gaussian peak into kink-anti-kink pair.}
    \label{fig:intro1}
}
\end{figure}

It is easy to construct a simple CCM for an oscillon that captures some of its key properties. In $\phi^4$ theory, if we use the background
\begin{equation}
\phi_{\rm bkg} = -1 + A(t)\Exp{-x^2/R(t)^2}\,,
\end{equation}
we get an effective Lagrangian
\begin{multline}\label{eq:osclag}
\sqrt{\frac{2}{\pi}}L_{\rm osc} = \frac{1}{2}R {\dot A}^2+\frac{1}{2} A\dot A \dot R +\frac{3{\dot R}^2 A^2}{8R} \\
-\frac{A^2}{2R}-2A^2 R +2\sqrt{\frac{2}{3}}A^3R -\frac{A^4R}{2\sqrt{2}}\,.
\end{multline}
Numerical investigations of this CCM\footnote{See Sec.~\ref{sec:IV} where we study essentially the same model.} reveal that for most initial values there exists, after a short initial phase, long quasi-periodic regime after which the `bubble' collapses exponentially fast ($A\to 0$ and $R\to \infty$). In that regard, the above CCM is qualitatively faithful to what is going on in field theory, at least for small energies. For energies large enough, we can expect that the oscillon decays into a $K\bar K$ pair (that can also undergo a few bounces before separation) as Fig.~\ref{fig:intro1} illustrates. 

This behavior is, of course, impossible to predict using \refer{eq:osclag}, which is inadequate for any situations where the field significantly departs from the Gaussian shape. Of course, one can attempt to remedy this by introducing more complicated ansatzes. 

Concrete CCMs will have limited applicability specific for a given situation. However, there are also general limitations. The most apparent one is pragmatical. Typically, when the number of collective coordinates exceeds $N=2$, the resulting formulae algebraically explode. 
Indeed, it is hard to come up with an $N$-dimensional moduli space for which the effective Lagrangian can be written down in a closed (and managable) form. We can see that for $K\bar K$ scattering CCMs which includes either normal modes \cite{2019arXiv190903128K} or higher Derrick's modes \cite{Adam:2021gat}. Simply put, $N$-point CCMs are typically \emph{algebraically intractable}.

This leads to a closely related limitation, namely that CCMs are usually not introduced as members of a perturbative schema that approaches full configuration space in a limit. In other words, CCMs are not usually \emph{exhaustive}. Having constructed one CCM, however successful, it is not apriori clear how to make a next step that would further deepen our understanding. 
In other words, typical CCMs are isolated islands of limited applicability that have no relation with one another.

Of course, this aspect might be both strength and weakness of the adiabatic approach. We can use it to zoom in on a particular aspect of the theory, such as $K\bar K$ scattering, and ignore everything else. By construction, a CCM reflects our \emph{apriori} knowledge about the solution space. Thus, it is difficult to be surprised and we cannot use CCMs to discover completely new phenomena.\footnote{Let us, however, point out that the spectral wall phenomenon \cite{Adam:2019xuc} might be a good counterexample for this generalizing statement.} 

In contrast, in this paper, we present a general-purpose CCM which attempts to overcome the aforementioned limitations. The three criteria that we demand are i) \emph{algebraic tractability} for an $N$-point CCM, ii) \emph{exhaustiveness}, e.g. that the CCM approaches the field theory as $N\to \infty$, and iii) \emph{qualitative} agreement between dynamical phenomena. Note that we do not (yet) demand \emph{quantitative} agreement but rather we regard it as an ultimate test. 

In particular, in this paper, we present \emph{simplest} (that we can imagine) general-purpose CCM that passes these criteria. However, we do not claim that it is also an optimal approach; indeed, there are deficiencies that we will point out. We regard our construction as a proof of concept.

We call our approach \emph{mechanization}; we replace a smooth field $\phi(x,t)$ by piece-wise linear segments -- i.e. a linear graph with a given number of `joints' (vertices) connected with straight `segments' (edges). The positions of joints and the associated field values (populations on the graph) are the time-dependent CCs. In other words, the continuous field -- upon mechanization -- becomes a jagged line resembling an axle connecting wheels on a steam train (see Fig.~\ref{fig:one}).  

Despite conceptual simplicity, the mechanization is surprisingly successful in capturing essential phenomena of scalar field theories, such as bouncing and bion pair-production  in $K\bar K$ scattering. As it does not require any prior knowledge, we can in fact \emph{discover} these phenomena by gradual exploration of `mech-models'. Indeed, the first two smallest values of $N$ reveals the mech-kink solution (a mechanical analog of a field-theoretical kink) and the mech-oscillon (an analog of oscillon) with the same CCMs as those in  \cite{RiceMJ} and Eq.~\refer{eq:osclag} without any prior insight. Furthermore, bouncing phenomenon, bion pair-production as well as more intricate patterns can be easily spotted in decay states of higher $N$ mech-oscillons. As these `mech-phenomena' gradually appear in an ordered sequence in complexity, we can use mechanization as a tool for \emph{systematic} exploration of a scalar field theory. 

The requirement for algebraic tractability is essential to this feature. 
As we will show, mechanization yields compact formulae for an arbitrary number of joints. In particular, we can write down the effective Lagrangian, the equations of motion and the metric for $N$ joints explicitly. 

In Sec.~\ref{sec:II} we describe the mechanization method, the effective Lagrangian, its symmetries, and the metric.
In Sec.~\ref{sec:III} we investigate mech-models with topological boundary conditions -- mech-kinks, while in Sec.~\ref{sec:IV} we turn our attention to topologically trivial configurations -- mech-oscillons.
Lastly, in Sec.~\ref{sec:V} we summarize our findings and discuss limitations and possible improvements.

\section{Mechanization}
\label{sec:II}

\subsection{Mech-field}

\begin{figure*}[htb!]
\begin{center}
\includegraphics[width=0.9\textwidth]{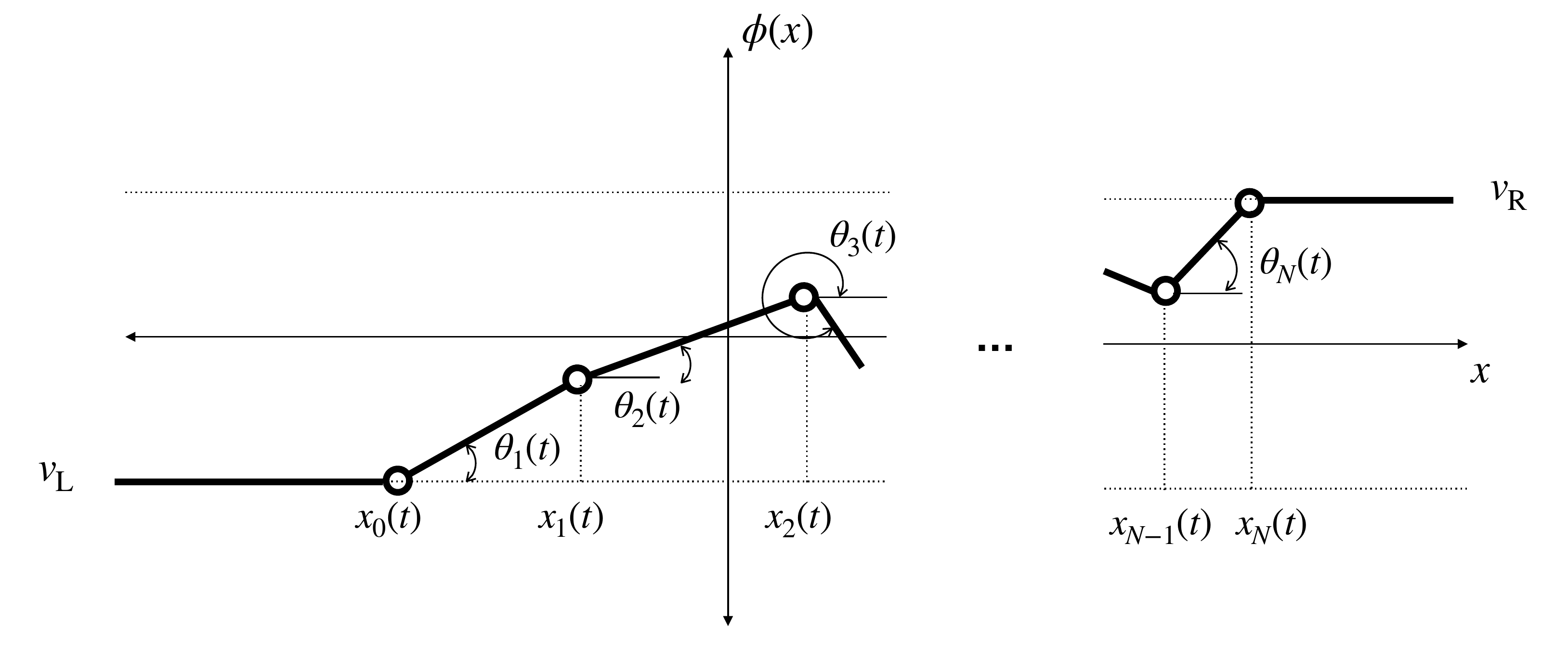}
\end{center}
{
    \caption{\small A depiction of a mech-field $\phi_M(x,t)$ as a sequence of $N$ straight segments connected via free-moving joints.}
    \label{fig:one}
}
\end{figure*}

\emph{Mechanization} is simply a replacement of a continuous field $\phi(x,t)$ with a piece-wise linear function which we dub a \emph{mech-field} $\phi_M(x,t)$. We express it mathematically as\footnote{We use the symbol $\xrightarrow[M]{}$ to indicate the mechanization of a continuous variable.} 
\begin{equation}\label{eq:decons2}
\phi(x,t) \xrightarrow[M]{} \phi_M(x,t) \equiv \sum\limits_{a=-1}^{N}\bigl(k_{a+1}(t)(x-x_a(t))+\phi_a(t)\bigr)\chi_a\,,
\end{equation}
where $x_a(t)$ are positions of $N+1$ joints, $\phi_a(t)\equiv \phi(x_a(t))$ are the field values and 
\begin{equation}
k_{a+1} \equiv \frac{\Delta \phi_a}{\Delta x_a} = \frac{\phi_{a+1}-\phi_a}{x_{a+1}-x_a} = \tan\bigl(\theta_a(t)\bigr)
\end{equation}
are the slopes of $a$-th segments (see Fig.~\ref{fig:one}). For our definition of mech-field to work, we formally assign joints to both spatial infinities, namely $x_{-1} = -\infty$ and $x_{N+1} = +\infty$. Furthermore, we always assume that the outermost segments are horizontal and lie in some vacua, i.e. $k_0 = k_{N+1} = 0$ and  $\phi_{-1} = \phi_0 = v_{\rm L}$, $\phi_{N}= \phi_{N+1} =  v_{\rm R} $, where $v_{\rm L(R)}$ are left (right) vacuum values. Also, the positions of segments form an  ordered sequence $x_0(t) \leq x_1(t) \leq \ldots \leq x_{N}(t)$.

The $\chi_a$'s are the indicator functions for each segment, namely $\chi_a = 1$ inside the interval $(x_a, x_{a+1})$ and zero outside:
\begin{equation}
\chi_a \equiv \theta(x-x_a)-\theta(x-x_{a+1}) = -\Delta \theta(x-x_a)\,,
\end{equation}
where the Heaviside step function is defined by
\begin{equation}\label{eq:theta}
\theta(x) = \lim_{\varepsilon \to 0^{+}}\int\limits_{-\infty}^{\infty}\frac{\diff \omega}{2\pi \iunit}\frac{\Exp{\iunit \omega x}}{\omega -\iunit \varepsilon} = \left\{\begin{matrix} 1 & \mbox{ if } x> 0  \\0 & \mbox{ if } x<0\end{matrix}\right.
\end{equation}
The value $\theta(0)$ is left unspecified, but it has no impact on physics. 

In certain sense, it might seem that there are two many variables. A standard way of discretization of space would be to replace a continuum variable $x$ with a lattice or in general with a fixed grid. In contrast, in our approach, the grid is itself dynamic. A dynamic grid is often employed to improve the precision of numerical integration of differential equations. There, however, the choice of grid spacing is dictated by optimization, while in our case the dynamics originate by construction.

\subsection{From mech-field to field}

It is easy to imagine a formal limit that give us back the original continuous field variable $\phi(x,t)$. If we send $N\to \infty$ in such a way that the joints become dense on the entire $x$-axis, i.e. $x_a(t) \to \xi$, the summation becomes an integration:
{\small \begin{gather}
 \phi_M(x,t) =  \sum\limits_{a=-1}^{N}\bigl(k_{a+1}(t)(x-x_a(t))+\phi_a(t)\bigr)\chi_a \nonumber \\
  \xrightarrow[N\to \infty]{}   \int\limits_{-\infty}^{\infty}\diff\xi\, \bigl(\phi^\prime(\xi)(x-\xi)+\phi(\xi)\bigr)\delta(x-\xi) = \phi(x)\,,
 \end{gather}}
 where we used the following correspondence: $k_{a+1} \to \phi^\prime(\xi)$ and $\chi_a \to \delta(x-\xi)\diff \xi$ as $N \to \infty$.

Even without taking the continuous limit, we can directly relate the discrete variables $\{x_a(t), \phi_a(t)\}$ and the time-dependent Fourier coefficients of $\phi(x,t)$ through Eq.~\refer{eq:theta}. In particular, after some algebra, we have
{\small \begin{equation}
\phi(x,t) \xrightarrow[M]{} \frac{1}{2}\int\limits_{-\infty}^{\infty}\frac{\diff \omega}{2\pi}\bigl(Y(\omega)\Exp{\iunit \omega x}+Y^{*}(\omega)\Exp{-\iunit \omega x}\bigr)+\frac{v_{\rm L}+v_{\rm R}}{2}\,,
\end{equation}}
where we denoted 
\begin{equation}\label{eq:fourier}
Y(\omega) \equiv -\frac{1}{\omega^2}\sum\limits_a^{N}\Delta k_a(t)\Exp{-\iunit \omega x_a(t)}\,.
\end{equation}
Notice that $Y(\omega)$ are nothing but Fourier coefficients of the mech-field $\phi_M(x,t)-v_{\rm L}$, while $Y^{*}(\omega)$ are Fourier coefficients of $\phi_M(x,t)-v_{\rm R}$.

It is worth mentioning that one can assign a set  of $2N$ discrete variables $\{x_a(t), \phi_a(t)\}$ to a given function  $\phi(x,t)$. This can be achieved by expanding $Y(\omega)$ into a Taylor series in $\omega$ up to $2N$-th power and matching the result with equivalent series for the Fourier image of $\phi(x,t) -v_{\rm L}$. From this point of view, mechanization can be roughly understood as an attempt to faithfully represent small frequencies (long wavelengths)  using the most simple functions available -- piece-wise linear functions.

\subsection{Effective Lagrangian}

We derive the effective Lagrangian by plugging the formula \refer{eq:decons2} into a canonical Lagrangian for a single scalar field
\begin{equation}\label{eq:contlag}
{\mathcal L} = \frac{1}{2}{\dot \phi}^2-\frac{1}{2}\phi^{\prime\, 2}-V(\phi)\,,
\end{equation}
where the potential has two degenerate minima $v_{\rm L}$ and $v_{\rm R}$, i.e. $V(v_{\rm L, R}) = V^{\prime}(v_{\rm L,R}) = 0$. 

The integration procedure is straightforward and yields compact formulas.  Let us illustrate it on  $\phi_M^{\prime\, 2}/2$. If we adopt the derivative $\phi_M^\prime$ as a differential consequence of \refer{eq:decons2}
\begin{gather}
\phi^\prime(x,t) \xrightarrow[M]{} \sum\limits_{a=-1}^{N}k_{a+1}\chi_a \equiv \phi_M^\prime (x,t)\,,
\end{gather}
it follows that
\begin{align}
\frac{1}{2}\lineint x\, \phi^{\prime\, 2} & \xrightarrow[M]{} \frac{1}{2}\sum\limits_{a,b}k_{a+1}k_{b+1} \lineint x\, \chi_a \chi_b \nonumber \\
& =\frac{1}{2}\sum\limits_{a=0}^{N-1} k_{a+1}^2 (x_{a+1}-x_a)\,,
\end{align}
where we have used an obvious identity $\chi_a \chi_b = \chi_a \delta_{ab}$. Notice that the resulting formula only involves the coupling of neighboring joints, and it is also manifestly positive (as $x_{a+1}\geq x_a$ by definition). 

Similarly, we calculate the kinetic energy 
\begin{equation*}
\frac{1}{2}\lineint x\, \dot \phi^2 \xrightarrow[M]{} \frac{1}{6}\sum\limits_{a=-1}^{N}\Delta x_a\Bigl(\Phi_{a+1,a}^{2}+\Phi_{a+1,a}\Phi_{a,a}+\Phi_{a,a}^2\Bigr)\,,
\end{equation*}
where we denoted $\Phi_{a,b}\equiv \dot \phi_{a}-k_{b+1}\dot x_{a}$ for convenience. The above expression can be recast in manifestly positive form as
\begin{multline}
\frac{1}{2}\lineint x\, \dot \phi^2 \xrightarrow[M]{} \frac{1}{12}\sum\limits_{a=0}^{N-1} \Delta x_a\Bigl(\bigl(\Phi_{a+1,a}+\Phi_{a,a}\bigr)^2   \\ +\Phi_{a+1,a}^2+\Phi_{a,a}^2\Bigr)\,,
\end{multline}
while most compact explicit formula reads
\begin{multline}
\frac{1}{2}\lineint x\, \dot \phi^2 \xrightarrow[M]{} \frac{1}{6}\sum\limits_{a=0}^{N-1}\dot k_{a+1}^2 \bigl(\Delta x_a\bigr)^3 
\\ +\frac{1}{2}\sum\limits_{a=0}^{N-1}\Delta x_a \bigl(\dot \phi_{a+1}-k_{a+1}\dot x_{a+1}\bigr)\bigl(\dot \phi_{a}-k_{a+1}\dot x_{a}\bigr)\,.
\end{multline}

Lastly, the potential energy is calculated as:
\begin{align}
\lineint x\, V(\phi) & \xrightarrow[M]{}  \lineint x\, V\Bigl(\sum\limits_{a=-1}^{N}\bigl(k_{a+1}(x-x_a)+\phi_a\bigr)\chi_a\Bigr)
\nonumber \\ & = \sum\limits_{a=-1}^{N} \int\limits_{x_a}^{x_{a+1}}\diff x\, V\Bigl(k_{a+1}(x-x_a)+\phi_a\Bigr) \nonumber \\
& = \sum\limits_{a=-1}^{N}\frac{1}{k_{a+1}}\int\limits_{\phi_a}^{\phi_{a+1}} \diff \xi \, V(\xi) \nonumber \\
& = \sum\limits_{a=0}^{N-1}\Delta x_a\frac{{\mathcal V}\bigl(\phi_{a+1}\bigr)-{\mathcal V}\bigl(\phi_{a}\bigr)}{\phi_{a+1}-\phi_a}\,,
\end{align}
where ${\mathcal V}$ is the primitive function of $V$.  
Notice that this is, indeed, positive: ${\mathcal V}$ is an increasing function and hence ${\mathcal V}\bigl(\phi_{a+1}\bigr)-{\mathcal V}\bigl(\phi_{a}\bigr)$ will be either positive or negative if $\phi_{a+1}-\phi_a$ is positive or negative.

Combining all our results, we can write the `mech-Lagrangian' compactly as
{\small \begin{gather}
 \lineint x\, {\mathcal L} \xrightarrow[M]{} \  L_{M}\,. \nonumber \\
L_{M} =  \sum\limits_{a=0}^{N-1}\Delta x_a\biggl[ \frac{1}{6}\Bigl(\Delta\dot \phi_a-\frac{\Delta \dot x_a}{\Delta x_a}\Delta\phi_a \Bigr)^2  -\frac{\Delta \phi_a^2}{2\Delta x_a^2}
\nonumber \\
+\frac{1}{2} \Bigl(\dot \phi_{a+1}-\frac{\Delta \phi_a}{\Delta x_a}\dot x_{a+1}\Bigr)\Bigl(\dot \phi_{a}-\frac{\Delta \phi_a}{\Delta x_a}\dot x_{a}\Bigr)
 - \frac{\Delta{\mathcal V}\bigl(\phi_{a}\bigr)}{\Delta \phi_{a}} \biggr]\,.
 \label{eq:efflag}
\end{gather} }

\subsection{Symmetries}

Let us consider the relation between symmetries of the field theory \refer{eq:contlag} and the mech-model \refer{eq:efflag}. 

First, the conserved quantities related to translational invariance of spacetime, i.e. $x^\mu \to x^\mu +a^\mu$ -- energy and momentum --  are preserved under mechanization. If we mechanize the continuous formulae we obtain the discrete momentum
 \begin{align}
P & = -\lineint x\, \dot \phi \phi^\prime  \xrightarrow[M]{}
\sum\limits_{a=0}^{N-1}\Delta \phi_a \bar U_a \equiv \sum\limits_{a=0}^{N-1}P_a \equiv P_{M}
\end{align}
and the discrete energy as
\begin{align}
E & = \lineint x\, \Bigl(\frac{1}{2}\dot \phi^2 +\frac{1}{2}\phi^{\prime\, 2} + V(\phi)\Bigr)\nonumber \\
&   \xrightarrow[M]{}
\sum\limits_{a=0}^{N-1} \biggl(\frac{1}{24}\dot k_{a+1}^2 \bigl(\Delta x_a\bigr)^3
 +\frac{1}{2}\bar U_a^2 \Delta x_a + \frac{1}{2}k_{a+1}^2 \Delta x_a  \nonumber \\ 
 & + \Delta x_a\frac{{\mathcal V}(\phi_{a+1})-{\mathcal V}(\phi_{a})}{\phi_{a+1}-\phi_a}\biggr) \equiv E_{M}\,,
\end{align}
where we introduced a quantity
\begin{align}
 \bar U_a & \equiv -\frac{1}{\Delta x_a}\int\limits_{x_a}^{x_{a+1}}\diff x\, \dot \phi =-\frac{1}{2}\Bigl(\Phi_{a+1,a}+\Phi_{a,a}\Bigr)\,,
\end{align}
which we loosely interpret as the average velocity of a segment.
For future reference, let us  also introduce the notion of segment's mass, namely twice of the static free energy:
\begin{equation}
M_a \equiv  \int\limits_{x_a}^{x_{a+1}}\diff x\, \phi^{\prime\, 2} = k_{a+1}^2\Delta x_a\,.
\end{equation} 
The total mass of the mech-field is thus 
\begin{equation}
M \equiv \sum\limits_{a=1}^{N-1}M_a = \sum\limits_{a=1}^{N-1} k_{a+1}^2\Delta x_a\,.
\end{equation}

As stated above, both momentum $P_M$ and energy $E_M$ are conserved quantities. Indeed, the Lagrangian $L_M$ \refer{eq:efflag} has a translational symmetry $x_a(t)\to x_a(t) +\alpha$ and it is also invariant under the time shift $t\to t+\alpha$. Invoking N\"other's theorem, we can compute the same expressions $P_M$ and $E_M$ directly from $L_{M}$. 

Unsurprisingly, the Lorentz invariance is generally lost. It is well known that invariance under boosts leads to a conserved  quantity\footnote{This follows from the conservation law $\partial_\mu {\mathcal M}^{\mu\nu\rho}=0$, where 
\begin{equation*}
{\mathcal M}^{\mu\nu\rho} = {\mathcal T}^{\mu\nu}x^\rho-{\mathcal T}^{\mu\rho}x^\nu\,,
\end{equation*}
with ${\mathcal T}^{\mu\nu}$ being the canonical energy-momentum tensor.
} 
\begin{equation}
J^{01}= - t P+\int\limits_{-\infty}^{\infty}\diff x\,  {\mathcal E}x\,, 
\end{equation}
that represents the uniform motion of the center of mass. A mechanized analog of this would generally not be a constant of motion.

One may expect that the Galilean boost should still be a good symmetry of $L_M$. However, if we make the transformation $x_a(t) \to x_a(t)- V t$ we find that the Lagrangian changes:
\begin{equation}
L_M \to L_M -P_M V + \frac{1}{2}M V^2\,.
\end{equation}
The problem is that the total mass $M$ is generally not a constant of motion, hence Galilean relativity is lost.

However, we can still relate solutions with different momenta. To do this, let us redefine the variables as
\begin{equation}
x_a(t) = \tilde x_a(t)+ r(t)\,,
\end{equation}
where $r(t)$ is a site-independent variable proportional to the average position (imposing the condition $\sum_a \tilde x_a = 0$). It follows that
\begin{align}
L_M(x) & = L_M(\tilde x) + P_M(\tilde x)\dot r+\frac{1}{2} \dot r^2 M(\tilde x)\,, \\
P_M(x) & = P_M(\tilde x) +M(\tilde x) \dot r\,,
\end{align} 
where we shown explicit dependence on $x = \{x_a\}$ or $\tilde x = \{\tilde x_a\}$ variables. We may easily identify equation of motion for $r(t)$ to be
\begin{equation}
\frac{\diff }{\diff t}\Bigl(M(\tilde x) \dot r +P_M(\tilde x)\Bigr) = 0\,.
\end{equation}
The expression in the brackets is a constant of motion, namely $P_M(x)$. Thus, if we find a solution $\tilde x_a(t)$ with momentum $P_M(\tilde x) \equiv \tilde P$, we can switch to a different solution with  momentum $P_M(x) \equiv P$ via
\begin{equation}
x_a(t) = \tilde x_a(t) +(P-\tilde P) \int\frac{dt}{M(\tilde x)}\,,
\end{equation}
where the second term is simply $r(t)$ integrated from its equation of motion.

Lastly, for completeness, let us display the equations of motion:
\begin{widetext}
\begin{gather}\label{eq:eomNx}
\Delta \biggl[\frac{\diff}{\diff t}\biggl(\frac{1}{6}\bigr(\Delta x_a\bigr)^2{\dot k_{a+1}}^2-\frac{1}{2}\Delta \phi_a \Phi_{a+1,a}\biggr)+\frac{1}{6}\frac{\Delta \dot\phi_a^3}{\Delta \dot \phi_a}-\frac{1}{6}k_{a+1}^2\Bigl(\frac{\Delta \dot x_a^3}{\Delta \dot x_a}-3\Bigr)-\frac{\Delta {\mathcal V}(\phi_a)}{\Delta \phi_a}\biggr] = - \frac{\diff }{\diff t}P_a\,, \\
\Delta \biggl[\frac{\diff}{\diff t}\biggl(-\frac{1}{3}\bigr(\Delta x_a\bigr)^2 \dot k_{a+1}+\frac{1}{2}\Delta x_a \Phi_{a+1,a}\biggr)-\frac{1}{6}\bigl(2\Delta \dot \phi_a \Delta \dot x_a+3 \dot x_{a+1}\dot \phi_a +3\dot x_{a}\dot \phi_{a+1}\bigr)+\frac{1}{3}k_{a+1}\Bigl(\frac{\Delta \dot x_a^3}{\Delta \dot x_a}-3\Bigr)+\frac{\Delta x_a}{\Delta \phi_a}\frac{\Delta {\mathcal V}(\phi_a)}{\Delta \phi_a}\biggr] \nonumber \\ \label{eq:eomNphi}
=\frac{\diff}{\diff t}\Bigl(\Delta x_a \bar U_a\Bigr) +V(\phi_{a+1})\Delta \frac{\Delta x_a}{\Delta \phi_a}\,.
\end{gather}
\end{widetext}
Note that summation of the first line implies conservation of momentum, i.e. $\dot P =\sum \dot P_a = 0$.

\subsection{Moduli space}

Integrating the kinetic energy over the field where certain parameters (moduli) are promoted to time-dependent variables, i.e. $\phi\bigl(x,t; \{ {\bf X}(t)\}\bigr)$, yields a metric on the moduli space spanned by the coordinates $\{ {\bf X}(t)\}$.

Our variables are $N+1$ positions $\{x_a(t)\}$ and $N-1$ field values $\{\phi_a(t)\}$ (the two outermost ones are forced to lie on vacua), i.e.  $\{ {\bf X}(t)\} \equiv \{x_a(t), \phi_a(t)\}$ giving rise to $2N$-dimensional moduli space. 

Studying the metric is especially useful for understanding the singularities of the moduli space. Given our heuristic choice of coordinates, we have no right to expect that the moduli space will be geodetically complete. Indeed, it is not. However, unlike most $N$-point CCM's, we can write down our metric on the back of an envelope:\footnote{To avoid awkward formulae, we extend the summation ranges for $\phi$ coordinates, although $\dot \phi_0 = \dot \phi_N = 0$.}
\begin{multline}
\frac{1}{2}\lineint x\, \dot \phi_M^2 = \frac{1}{2} \dot {\bf X}^T {\bf g^{XX}}\dot {\bf X}
\\ = \frac{1}{2}\sum\limits_{a,b =0}^{N}\Bigl( g_{ab}^{xx}\dot x_a \dot x_b+2g_{ab}^{x\phi}\dot x_a \dot \phi_b + g_{ab}^{\phi\phi}\dot \phi_a\dot \phi_b\Bigr)\,,
\end{multline}
where
\begin{gather}
g_{aa}^{xx} = \frac{1}{3}\frac{(\Delta \phi_a)^2}{\Delta x_a}+ \frac{1}{3}\frac{(\Delta \phi_{a-1})^2}{\Delta x_{a-1}}\,, 
\\ g_{a\, a+1}^{xx} = \frac{1}{6}\frac{(\Delta \phi_a)^2}{\Delta x_a}\,,
\hspace{3mm} g_{a\, a-1}^{xx} = \frac{1}{6}\frac{(\Delta \phi_{a-1})^2}{\Delta x_{a-1}}\,, \\
g_{aa}^{x\phi} = -\frac{1}{3}\bigl(\phi_{a+1}-\phi_{a-1}\bigr)\,, \hspace{3mm}
g_{a\,a+1}^{x\phi} = -\frac{1}{6}\Delta \phi_a\,, \\
g_{a\,a-1}^{x\phi} = -\frac{1}{6}\Delta \phi_{a-1}\,, \hspace{3mm}
g_{aa}^{\phi\phi} = \frac{1}{3}\bigl(x_{a+1}-x_{a-1}\bigr)\, \\
g_{a\, a+1}^{\phi\phi} = \frac{1}{6}\Delta x_a\,, \hspace{3mm}
g_{a\, a-1}^{\phi\phi} = \frac{1}{6}\Delta x_{a-1}\,.
\end{gather}
The components $g^{xx}$, $g^{x\phi}$ and $g^{\phi\phi}$ are tri-diagonal due to only neighboring interactions.

There are many singularities in this metric, but it is quite straightforward to appreciate why.
Most apparently, when the $x$-distance of neighboring joints becomes zero, i.e. $\Delta x_a =0$, the components of $g^{xx}$ diverge. Indeed, when $\Delta x_a=0$ and  $\phi_a \not = \phi_{a+1}$ the mech-field becomes multi-valued.\footnote{Let us note that when $\Delta x_a \to 0$ and $\Delta \phi_a \to 0$, such that the ratio $(\Delta \phi_a)^2/\Delta x_a$ is kept fixed, the metric remains completely regular. However, this is not true for components of the Riemann tensor, which diverge.}
It is possible to remove these singularities to infinity by a change of coordinates, e.g.
\begin{equation}
x_a(t) = x_0(t)+ \sum\limits_{c=0}^{a}\Exp{b_c(t)}\,, 
\end{equation}
so that $\Delta x_a = \Exp{b_{a+1}}$. This also enforces the ordering $x_0 < x_1 < \ldots x_N$. 
We have used these coordinates in all our numerical calculations.

A more subtle issue arises when a segment becomes flat, i.e. $\Delta \phi_a = 0$. If the segment in question is not on the edges, then the metric remains well-defined. In fact, $g^{xx}$ and $g^{x\phi}$ become block-diagonal, suggesting dynamical decoupling of the left and right parts of the mech-field, in agreement with intuition. If the segment is on the border, i.e. $\Delta \phi_0 =0$ or $\Delta \phi_N =0$, this represents a sudden uncoupling of an edge joint. Both cases manifest an abrupt change in the mech-field, which is confirmed by looking at the formula for the determinant:
{\small \begin{equation}
\bigl|{\bf g^{XX}}\bigr| = \frac{1}{12^N}\prod\limits_{a=0}^{N-1}\frac{(\Delta \phi_a)^4(\Delta \phi_a \Delta x_{a+1}-\Delta \phi_{a+1}\Delta x_a)^2}{(\Delta x_a)^2}\,.
\end{equation}}
As we see,  the volume measure vanishes if any $\Delta \phi_a =0$.
 
We can also observe that when neighboring slopes become identical, i.e. $k_{a} = k_{a+1}$, the determinant vanishes too.  However, these instances seem to be coordinate singularities. We checked that for the first few $N$ the expression $\sqrt{{\bf g^{XX}}}{\bf R^{XX}}$, where ${\bf R^{XX}}$ is the Ricci scalar, is well defined in the limit of equal subsequent slopes.

\section{Mech-kink}
\label{sec:III}

In this  section, we shall begin a systematic investigation of solutions of mech-model \refer{eq:efflag} for increasing value of $N$. Here we mostly focus on static solutions of topological configurations, that we call mech-kinks. As we shall see, the model for a simplest mech-kink is totally integrable and formally equivalent to the relativistic CCM of a BPS kink \cite{Adam:2021gat}.

\subsection{$N=1$ mech-kink}

\begin{figure}
\begin{center}
\includegraphics[width=0.95\columnwidth]{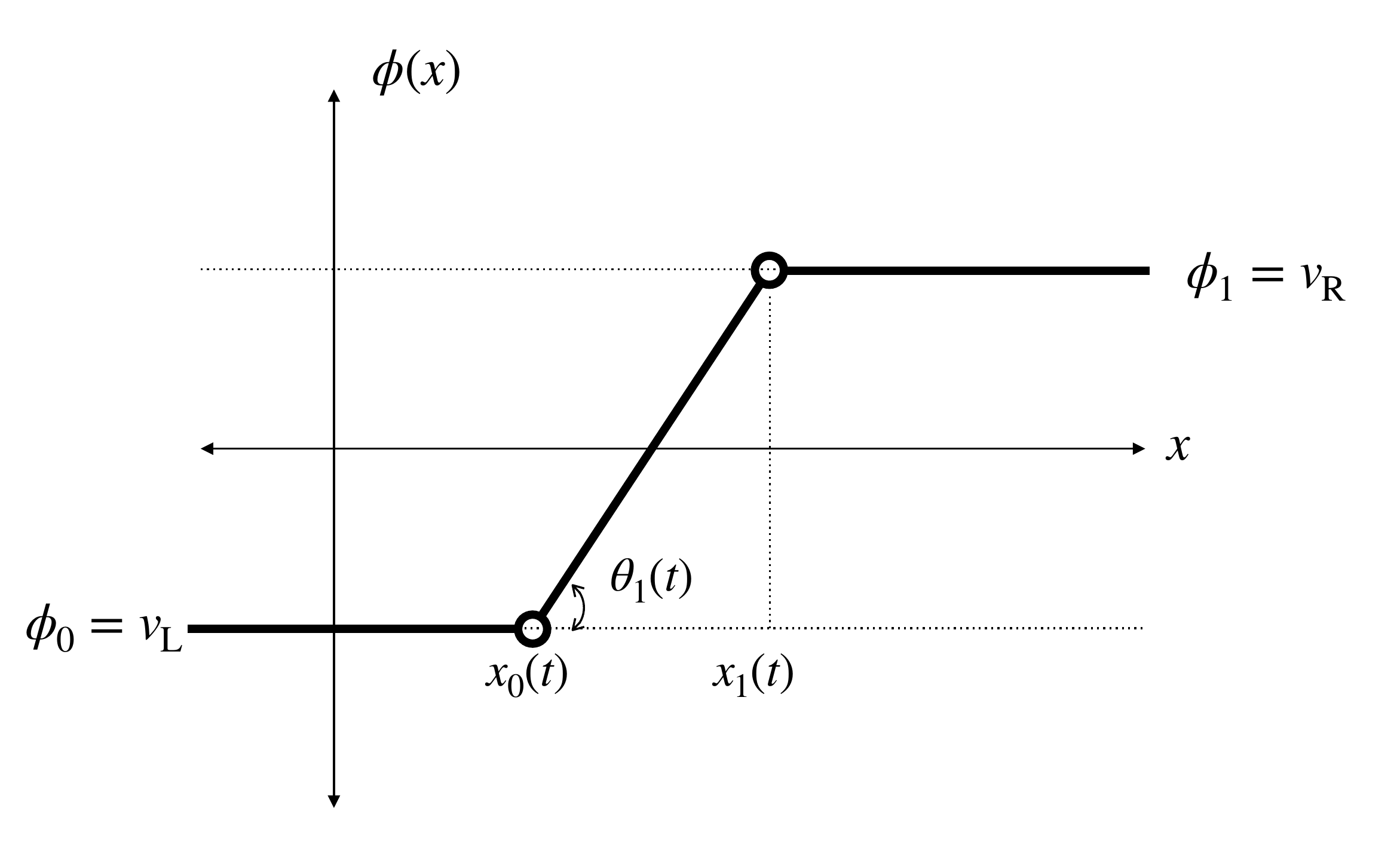}
\caption{\small Simplest mechanical model of a kink = `mech-kink'.}
\label{fig:two}
\end{center}
\end{figure}

Let us consider $N=1$ case with topological boundary conditions, i.e. $\phi_{0}= v_{\rm L} \not = \phi_1 = v_{\rm R}$, which we call a \emph{mech-kink} (see Fig.~\ref{fig:two}). The Lagrangian \refer{eq:efflag} for $N=1$ is explicitly given by 
\begin{equation}\label{eq:n1mechlag}
L_{M}^{N=1} = \frac{\bigl(\Delta \phi_0\bigr)^2}{24\Delta x_0} \Bigl(\bigl(\Delta \dot x_0\bigr)^2-12 + 12\, \dot {\bar x}^2\Bigr) - \kappa \Delta x_0\,,
\end{equation} 
where we have denoted the average position $\bar x \equiv (x_0+x_1)/2$ and the constant
\begin{equation}
\kappa \equiv \frac{1}{v_{\rm R}-v_{\rm L}}\int\limits_{v_{\rm L}}^{v_{\rm R}}\diff \xi\, V(\xi)\,.
\end{equation}

The equations of motion reads
{\small \begin{align}
\label{eq:eomk1} \frac{\diff}{\diff t} \Bigl(\frac{(\Delta \phi_0)^2 \Delta \dot x_0}{12 \Delta x_0}\Bigr)  & = -\frac{\bigl(\Delta \phi_0\bigr)^2}{24(\Delta x_0)^2} \Bigl(\bigl(\Delta \dot x_0\bigr)^2-12 + 12\, \dot {\bar x}^2\Bigr) - \kappa\,, \\
\label{eq:eomk2} \frac{\diff}{\diff t} \Bigl(\frac{(\Delta \phi_0)^2  \dot {\bar x}}{\Delta x_0}\Bigr)  & = 0\,.
\end{align}}
The two conserved quantities reads:
\begin{align}
E_{M}^{N=1}  &=  \frac{\bigl(\Delta \phi_0\bigr)^2}{24\Delta x_0} \Bigl(\bigl(\Delta \dot x_0\bigr)^2+12 + 12\, \dot {\bar x}^2\Bigr) +\kappa \Delta x_0\,, \\
P_{M}^{N=1} & = \frac{\bigl(\Delta \phi_0\bigr)^2}{\Delta x_0} \dot {\bar x}\,.
\end{align}
Note that $\Delta \phi_0 = v_{\rm R}-v_{\rm L}$ is a constant.

Let us first consider a static mech-kink: $\Delta \dot x_0 = \dot {\bar x} = 0$. 
Solving the equations of motion \refer{eq:eomk1}-\refer{eq:eomk2} we find the energy $E_{M,\, {\rm static}}^{N=1} \equiv m_K$ and the width $R_{K} \equiv \Delta x_0$ to be
\begin{equation}
m_K = (v_{\rm R}-v_{\rm L})\sqrt{2\kappa}\,, \hspace{5mm}
R_K =\frac{v_{\rm R}-v_{\rm L}}{\sqrt{2\kappa}}\,.  
\end{equation}
For concreteness, if we consider $\phi^4$ potential $V(\phi) =(1-\phi^2)^2/2$  we have $v_{\rm R} = -v_{\rm L} = 1$ and $\kappa = 4/15$. The corresponding numbers for a static mech-kink reads $m_K = \sqrt{32/15}$ and $R_K = \sqrt{15/2}$.
Notice that mech-kink's mass is equal to the total mass of the mech-field, i.e. $m_K = M_0 = (\Delta \phi_0)^2/\Delta x_0$.

Surprisingly, the Lagrangian $L_M^{N=1}$ is formally equivalent to a relativistic CCM for a BPS kink (Eq.~(II.17) of \cite{Adam:2021gat}):
\begin{equation}\label{eq:bpskinklag}
L[a,b] = \frac{1}{2}M b \dot a^2 +\frac{Q}{2b^3}\dot b^2-\frac{1}{2}M\Bigl(b+\frac{1}{b}\Bigr),
\end{equation}
which is obtained in $\phi^4$ theory using the anstaz $\phi_K = \tanh\bigl(b (x-a)\bigr)$. Here, $M=4/3$ is the kink's mass and $Q$ is the second moment of static kink's energy density:
\begin{equation}
Q = \int\limits_{-\infty}^\infty \diff x\, x^2\phi_K^{\prime\, 2}(x) = \frac{\pi^2-6}{9}\,.
\end{equation}

The similarity of \refer{eq:n1mechlag} and \refer{eq:bpskinklag} becomes explicit if we set
\begin{equation}
\Delta x_0 = \frac{R_K}{b}\,, \hspace{5mm} \bar x = a
\end{equation}
The Lagrangian $L_{M}^{N=1}$ switches to a form:
\begin{equation}\label{eq:bpsmechkinklag}
L_M[a,b] = \frac{1}{2}m_k b \dot a^2 +\frac{q_M}{2 b^3}\dot b^2-\frac{1}{2}m_k\Bigl(b+\frac{1}{b}\Bigr)\,,
\end{equation}
where 
\begin{equation}
q_M \equiv \int\limits_{-\infty}^{\infty}\diff x\, x^2 \phi_M^{\prime\, 2}(x) = \frac{1}{12}R_K\Delta \phi_0^2\,.
\end{equation}
For $\phi^4$ model $m_K = \sqrt{32/15} \approx 1.46$ which is not far off from the BPS value $4/3 \approx 1.33$. However, $q_M = \sqrt{5/6} \approx 0.91$ is more than twice of the field-theoretical value $Q \approx 0.43$. 

Let us stress that apriori we had no right to expect this formal correspondence. The mech-field $\phi_M^{N=1}$ is not a BPS solution of the field theory, yet we see \emph{precisely} the same effective Lagrangian as for the BPS kink.
There seems to be a certain universality of CCMs, regarding the structure of terms. Indeed, it is easy to show that the same effective Lagrangian as in \refer{eq:bpskinklag} arises for any background $\phi = f(b(x-a))$ provided that $f^{\prime\, 2}$ has finite first three moments. Thus, the key ingredient is not the shape of the solution but the inclusion of a scaling modulus. It is somewhat of an accident that $N=1$ mech-field also falls into this class of backgrounds, as the scaling modulus appears as the slope of the middle segment.

Given the exact correspondence with relativistic CCM, it is not surprising that we will find the same results, namely the Lorentz covariance and the existence of a Derrick mode. However, let us stress that from the point of view of the mech-model both results are very unexpected!

Indeed, if we consider a mech-kink moving with a uniform velocity $v$, i.e. $\Delta \dot x =0$, $\dot {\bar x} = v$, the equations of motion \refer{eq:eomk1}-\refer{eq:eomk2}  can be easily solved. We find that energy $E_M^{N=1}\equiv E_K$ and width $\Delta x_0$ follow formulae for a relativistic particle:
\begin{equation}
E_K = \frac{m_K}{\sqrt{1-v^2}}\,, \hspace{5mm} \Delta x_0  = R_K \sqrt{1-v^2} 
\end{equation}

Let us now fix the center position to the origin ${\bar x} = 0$. If the energy $E_K$ is above the static energy, i.e. $E_K >m_K$, the mech-kink oscillates with angular frequency $\omega_M = \sqrt{12}/R_K$
\begin{equation}\label{eq:oscmk}
\Delta x_0 = \frac{E_K}{2\kappa}-\frac{\sqrt{E_K^2-m_k^2}}{2\kappa}\sin \bigl(\omega_M t\bigr)\,.
\end{equation}  
This vibrational mode is independent on the shape of the potential. Thus, it is not a shape-mode -- a massive normal mode of the kink. Indeed, not all kinks have the shape mode as is well known \cite{Dorey:2011yw}. This mode is more appropriately identified with the so-called Derrick mode, which arises due to infinitesimal scaling of the static solution and which exists for all kinks \cite{Adam:2021gat}.

In the field theory, the frequency of the Derrick mode is $\omega_D^2 = M/Q$. In the mech-model, it is again formally the same $\omega_M^2 = m_K/q_M$. Specifically, in $\phi^4$ theory, $\omega_D^2 \approx 3.1$ and $\omega_M^2 \approx 4.4$.

We can construct a general solution of \refer{eq:eomk1}-\refer{eq:eomk2} as we have the same number of unknowns, namely $\Delta x_0$ and $\bar x$, and constants of motion, i.e. $E_K$ and $P_K$. Indeed, Eq.~\refer{eq:eomk2} is equivalent to $\dot P_K = 0$ which implies the conservation of the momentum. Furthermore, Eq.~\refer{eq:eomk2} can be linearized as
\begin{equation}
\Delta \ddot  x_0 = \frac{12 E_K}{(\Delta \phi_0)^2}-\frac{12}{(\Delta \phi_0)^2}\Delta x_0 \Bigl(\kappa + \frac{P_K^2}{(\Delta \phi_0)^2}\Bigr)\,.
\end{equation}
The general solution thus reads
\begin{multline}
\Delta x_0 = R_K \frac{m_K}{P_K^2 +m_K^2}\biggl(E_K \\
-\sqrt{E_K^2 -P_K^2 -m_K^2}\sin\Bigl(\frac{\sqrt{12}t \sqrt{P_K^2+m_K^2}}{m_K R_K}\Bigr)\biggr)\,.
\end{multline}

\subsection{$N>1$ static mech-kinks}

\begin{figure*}
\begin{center}
\includegraphics[width=0.95\textwidth]{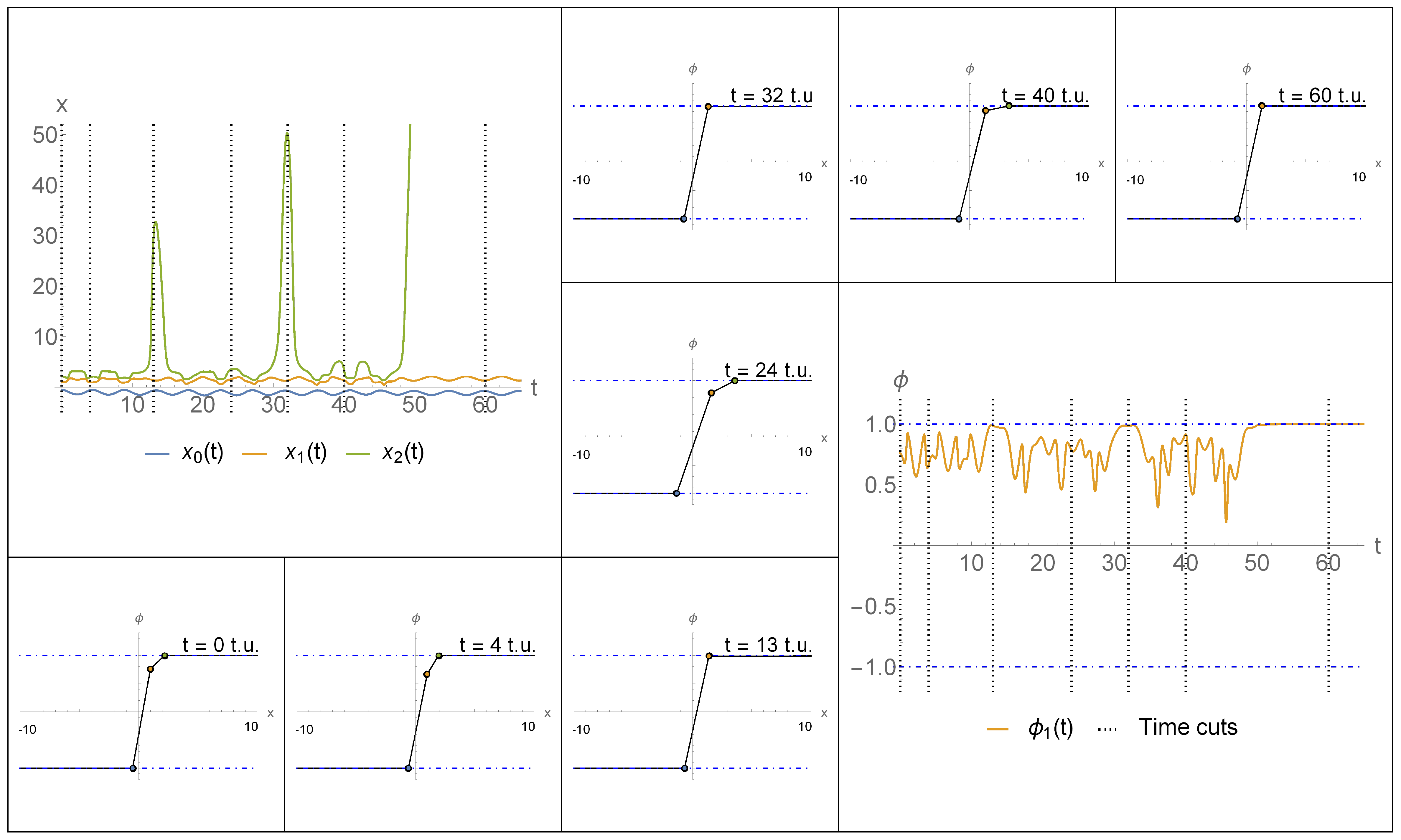}
\caption{\small An example of joint-ejection for $N=2$ mech-kink, where the third joint escapes to $\infty$.}
\label{fig:eight}
\end{center}
\end{figure*}

\begin{figure*}
\begin{center}
\includegraphics[width=0.95\textwidth]{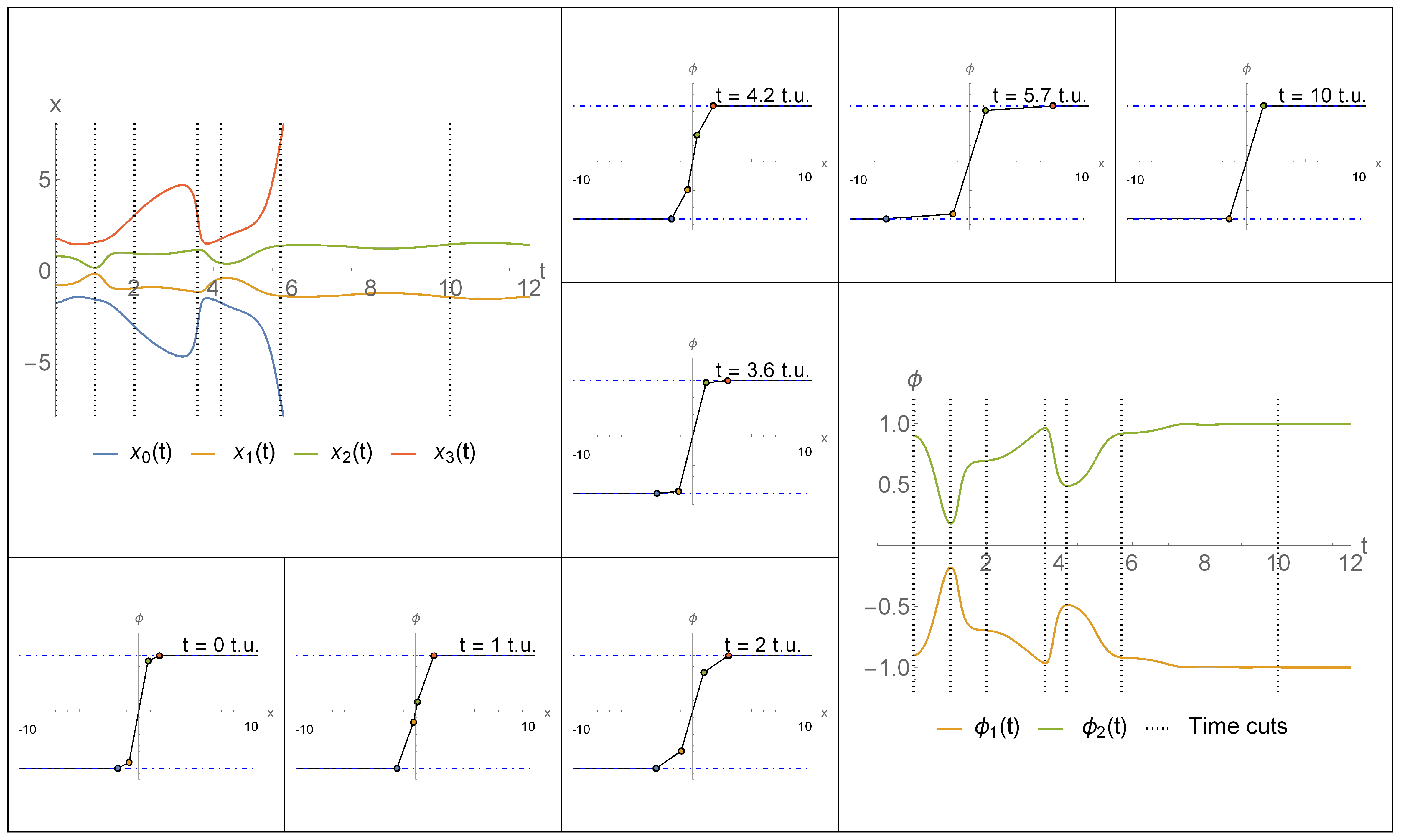}
\caption{\small A symmetric example where the outermost joints of $N=4$ mech-kink get ejected, leaving behind slightly excited $N=2$ mech-kink.}
\label{fig:nine}
\end{center}
\end{figure*}

\begin{figure}
\begin{center}
\includegraphics[width=0.95\columnwidth]{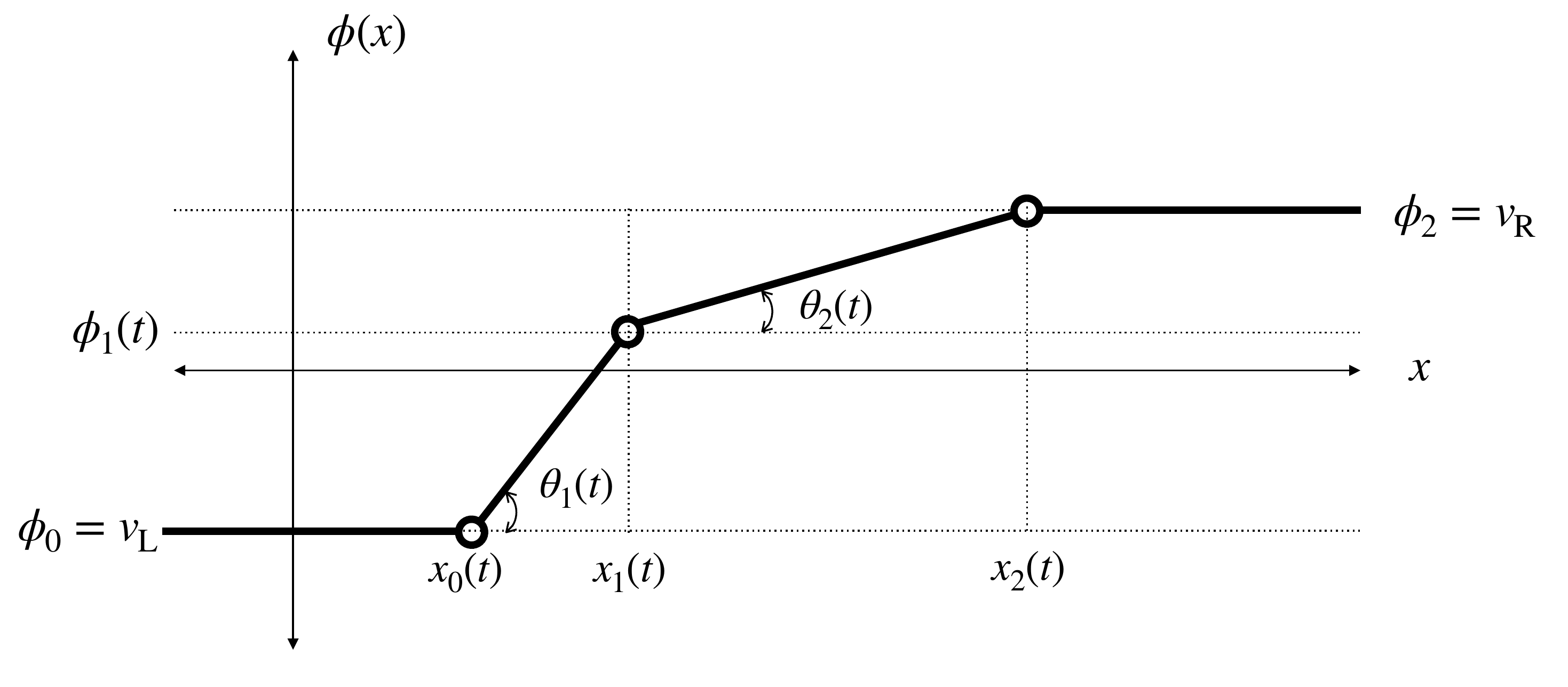}
\caption{\small Schematic depiction of $N=2$ mech-kink.}
\label{fig:three}
\end{center}
\end{figure}

All non-trivial static configurations are mech-kinks, by which we mean any configuration for which the outer segments lie in different vacua.\footnote{We exclude the possibility of an inner segment lying in a vacuum as that would lead to decoupled mech-kink and anti-mech-kink solutions. These solutions nevertheless exist and can be considered as exact free mech-kink gas solutions.}
For instance, we depict $N=2$ mech-kink on Fig.~\ref{fig:three}.

The static equations of motion translate to the following conditions for the lengths of the segments
\begin{align}
\Delta x_0 & = \sqrt{\frac{(\phi_1-v_{\rm L})^3}{2\bigl( {\mathcal V}(\phi_1)-{\mathcal V}(v_{\rm L})\bigr)}}\,,
\\
\Delta x_1 & = \sqrt{\frac{(v_{\rm R}-\phi_1)^3}{2\bigl( {\mathcal V}(v_{\rm R})-{\mathcal V}(\phi_1)\bigr)}}\,,
\end{align}
while the field values satisfy an algebraic equation
\begin{equation}
k_0 -k_1 + V(\phi_1)\biggl(\frac{1}{k_0}-\frac{1}{k_1}\biggr) = 0\,,
\end{equation}
where $k_0 = (\phi_1 - v_{\rm L})/\Delta x_0$ and $k_1 = (v_{\rm R}-\phi_1)/\Delta x_1$ are the slopes of the segments.
This condition can be reduced to a simpler form, namely $V(\phi_1) = k_0 k_1$, provided that $k_0 \not =k_1$. If $k_0 = k_1$ we would simply get back $N=1$ mech-kink solution.\footnote{This is true also for trivial solutions $\phi_1 = v_{\rm L}$ and $\phi_1 = v_{\rm R}$.}

\begin{figure}
\begin{center}
\includegraphics[width=0.95\columnwidth]{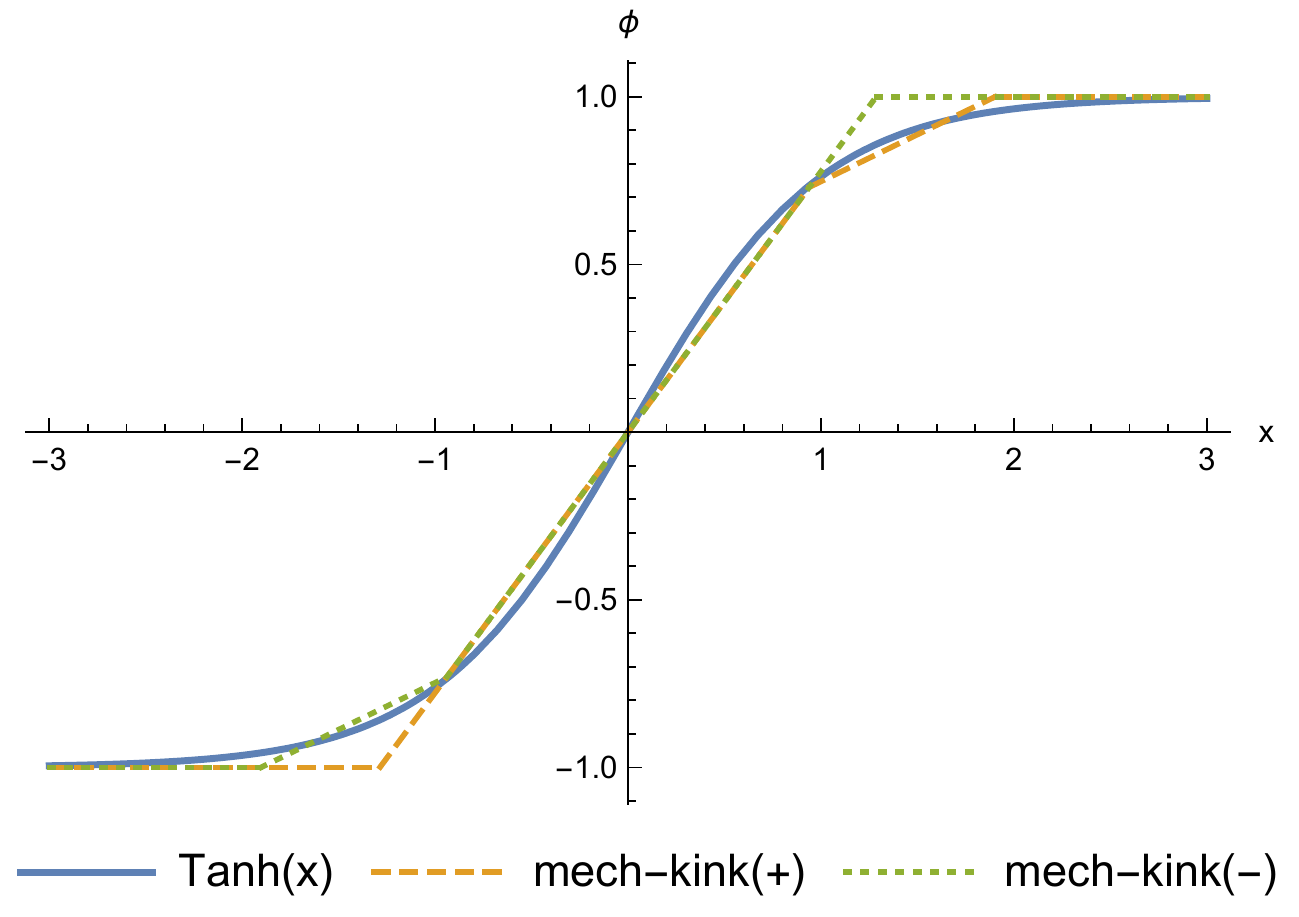}
\caption{\small Comparison between exact kink solution in the double-well model and two static solutions for $N=2$ mech-kinks.}
\label{fig:four}
\end{center}
\end{figure}

To be concrete, let us consider $\phi^4$ model. We found two minimum-energy solutions that are $\Delta x_0 \leftrightarrows \Delta x_1$, $\phi_1 \leftrightarrows -\phi_1$ reflections of each other (see Fig.~\ref{fig:four}):
\begin{align}
\phi_1 & = \frac{\pm 1}{12}\sqrt{139 -\sqrt{3865}} \approx \pm 0.73\,, \\
\Delta x_0 & = \frac{1}{80}\Bigl(3\sqrt{1373+7\sqrt{3865}}\pm \sqrt{3\bigl(4519-59\sqrt{3865}\bigr)}\Bigr) \nonumber \\
& \approx 1.594\pm 0.631\,, \\
\Delta x_1 & = \frac{1}{80}\Bigl(3\sqrt{1373+7\sqrt{3865}}\mp \sqrt{3\bigl(4519-59\sqrt{3865}\bigr)}\Bigr)\nonumber \\ 
& \approx 1.594\mp 0.631\,.
\end{align}
Their energies are the same
\begin{multline}
E_{K,\pm} = \frac{1}{216}\sqrt{\frac{1}{10}(989543-773\sqrt{3865})} \approx 1.42 \\ < E_K = \sqrt{\frac{32}{15}} \approx 1.46
\end{multline}
and are roughly 3\% smaller than the energy of $N=1$ mech-kink. This is understandable as adding more segments should get us closer to the exact kink mass $M= 4/3 \approx 1.33$.

\begin{figure}
\begin{center}
\includegraphics[width=0.95\columnwidth]{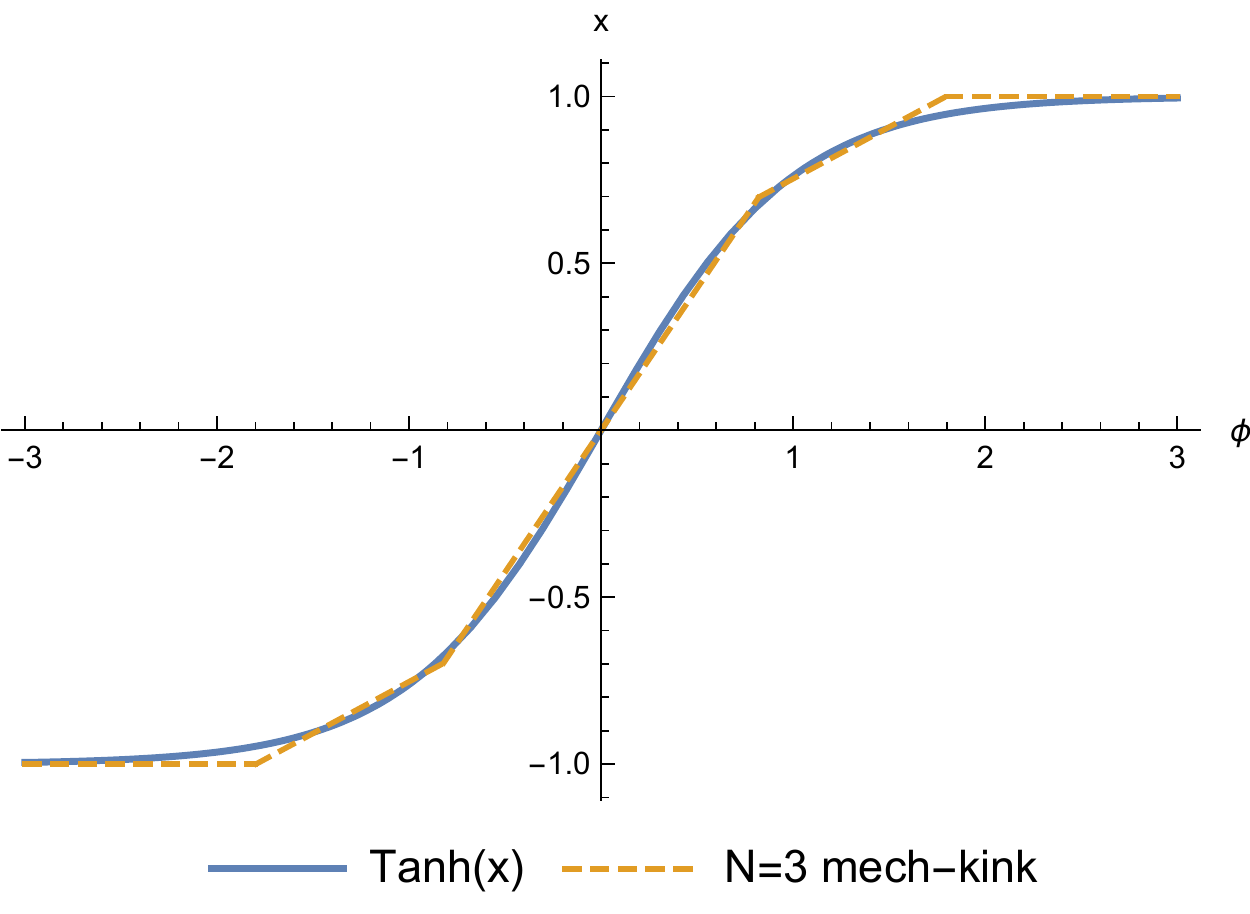}
\caption{\small Comparison between the kink solution in the double-well model and a static $N=3$ mech-kink.}
\label{fig:five}
\end{center}
\end{figure}

For $N=3$, there is  only a single minimum-energy solution which we show in Fig.~\ref{fig:five}. Its energy is $E_K^{N=3} \approx 1.37$.

For $N=4$, we find again two mirror-image solutions with energies $E_K^{N=4} \approx 1.36$. This pattern repeats. For odd $N$ we get a unique solution, while for even $N$ we find two degenerate solutions.

The static equations of motion (see Eqs.~\refer{eq:eomNx}-\refer{eq:eomNphi}) for arbitrary $N$ reduces to
\begin{gather}
k_{a+1} \equiv \frac{\Delta \phi_a}{\Delta x_a} = \sqrt{\frac{2\bigl({\mathcal V}(\phi_{a+1})-{\mathcal V}(\phi_{a})\bigr)}{\phi_{a+1}-\phi_a}}\,, \\
k_{a+1}k_a = 2 V(\phi_a)\,.
\end{gather}
Notice that in the continuous limit, the first formula becomes the BPS equation for a kink, i.e. $\phi^\prime = \sqrt{2V(\phi)}$.
We can simplify the above system to a set of $N-1$ algebraic equations for $\phi_a$'s, namely
{\small \begin{equation}
V(\phi_a) = \sqrt{\frac{\bigl({\mathcal V}(\phi_{a+1})-{\mathcal V}(\phi_{a})\bigr)\bigl({\mathcal V}(\phi_{a})-{\mathcal V}(\phi_{a-1})\bigr)}{\bigl(\phi_{a+1}-\phi_a\bigr)\bigl(\phi_{a}-\phi_{a-1}\bigr)}}\,.
\end{equation}}

It is also easy to see that every static solution has its boosted version, meaning that  the energy is equal to the static energy times the Lorentz $\gamma$ factor, the field values are unchanged, while $\Delta x_a$'s  are contracted by $\sqrt{1-v^2}$.

\subsection{Dynamics of mech-kinks}

There are several questions about mech-kinks that interest us which are, however, outside the scope of this paper. For instance, we would like to know how fast the static energy approaches the BPS bound as a function of $N$. 

\begin{figure*}
\begin{center}
\includegraphics[width=0.95\textwidth]{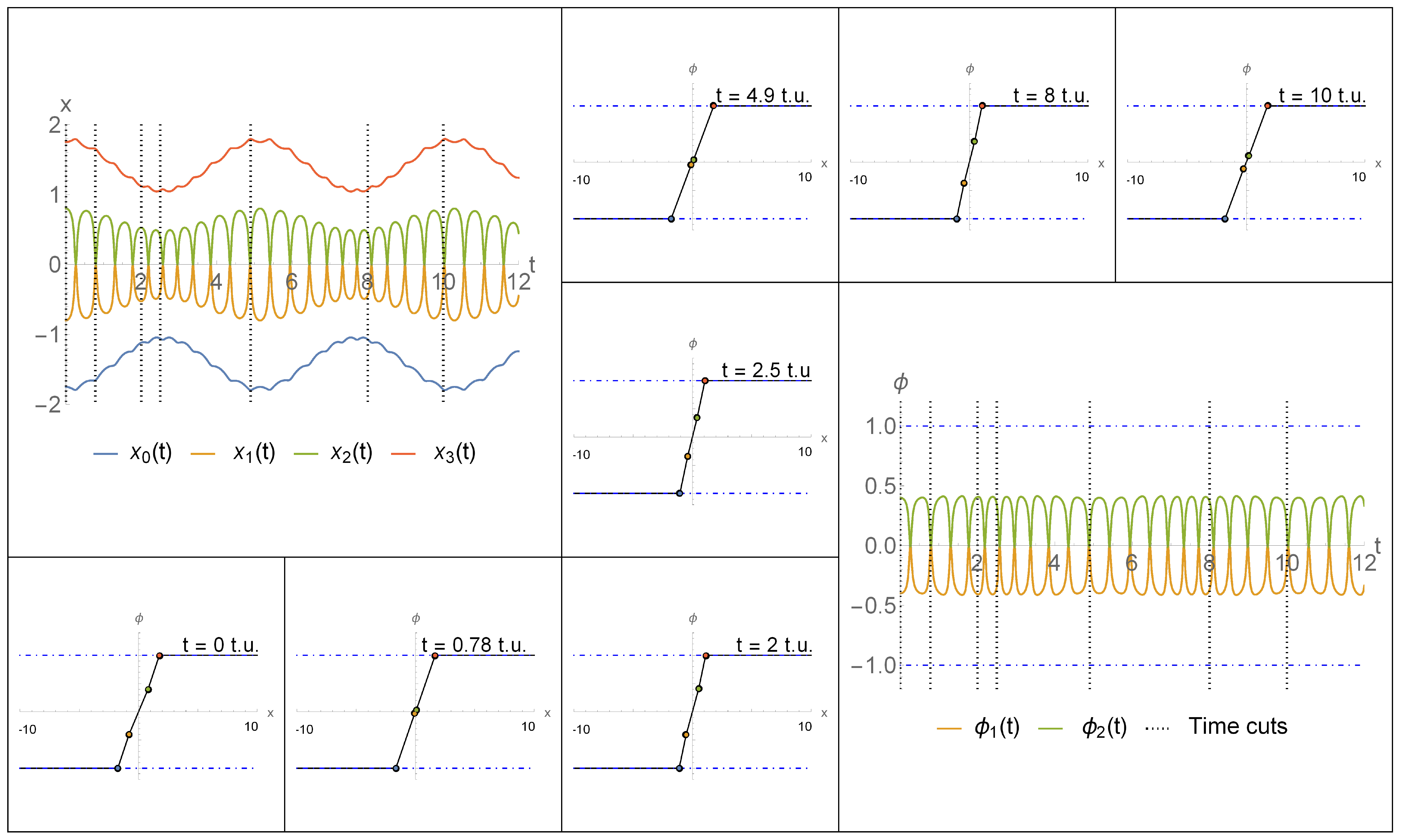}
\caption{\small Quasi-periodic oscillations in $N=4$ mech-kink.}
\label{fig:seven}
\end{center}
\end{figure*}

Regarding the dynamics, an important query for $N\geq 2$ mech-kinks is whether there exist any exact periodic solutions. We have not been able to find the answer analytically. Numerically, however, we have glimpsed a promising candidate (see Fig.~\ref{fig:seven}). 

\begin{figure*}
\begin{center}
\includegraphics[width=0.95\textwidth]{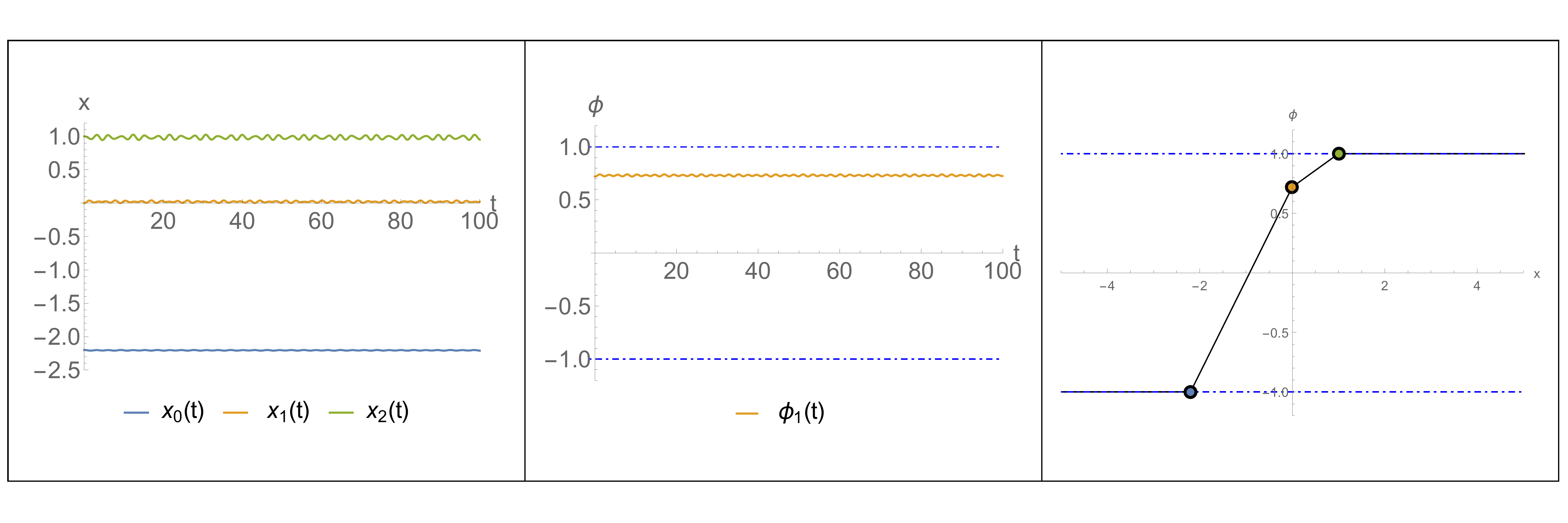}
\caption{\small Small perturbation of static $N=2$ mech-kink leading to quasi-periodic oscillations.}
\label{fig:six}
\end{center}
\end{figure*}

A related problem is the investigation of normal modes of the mech-kinks. In particular, we would like to study how the spectrum of small fluctuations varies with increasing $N$, how many spurious modes there are (compared with field theory), etc.  In Fig.~\ref{fig:six}, we show a numerical solution of a slightly perturbed static $N=2$ mech-kink indicating the presence of at least one normal mode.

Ultimately, we would like to categorize the dynamics of mech-kinks for a vast set of initial conditions to obtain a robust understanding of Cauchy's problem for each $N$. This task, however, is too time-consuming for our purpose here. At present, 
we have only sampled the evolution of a few $N\geq 2$ mech-kinks for random initial conditions via numerical integration of equations of motion. 

One phenomenon that we found to be endemic for all $N \geq 2$ mech-fields is the \emph{joint-ejection}, i.e. when one of the outer joints rapidly approaches vacuum and flies either to the left or right infinity, leaving behind an effective $N-1$ mech-field. We observe this already for $N=2$ mech-kinks, where the left-over piece is an excited $N=1$ mech-kink (see Fig.~\ref{fig:eight}). Curiously, two joint-ejections can happen simultaneously for the boundary joints, leaving behind $N-2$ mech-field (see Fig.~\ref{fig:nine}).

The joint-ejection is indicative of a general tendency for a mech-field to simplify itself as time increases. In fact, it is not unreasonable to think  that all initial configurations for mech-kinks eventually settle to an exited $N=1$ mech-kink.

%%%%%%%%%%%%%%%%%%%%%%%%%%%%%%%%%%%%%%%%%%%%%%%%%%%%
%%%%%%%%%%%%%%%%%%%%%%%%%%%%%%%%%%%%%%%%%%%%%%%%%%%%
%%%%%%%%%%%%%%%%%%%%%%%%%%%%%%%%%%%%%%%%%%%%%%%%%%%%
%%%%%%%%%%%%%%%%%%%%%%%%%%%%%%%%%%%%%%%%%%%%%%%%%%%%
\section{Mech-oscillon }
\label{sec:IV}
%%%%%%%%%%%%%%%%%%%%%%%%%%%%%%%%%%%%%%%%%%%%%%%%%%%%
%%%%%%%%%%%%%%%%%%%%%%%%%%%%%%%%%%%%%%%%%%%%%%%%%%%%
%%%%%%%%%%%%%%%%%%%%%%%%%%%%%%%%%%%%%%%%%%%%%%%%%%%%
%%%%%%%%%%%%%%%%%%%%%%%%%%%%%%%%%%%%%%%%%%%%%%%%%%%%

\begin{figure}
\begin{center}
\includegraphics[width=0.95\columnwidth]{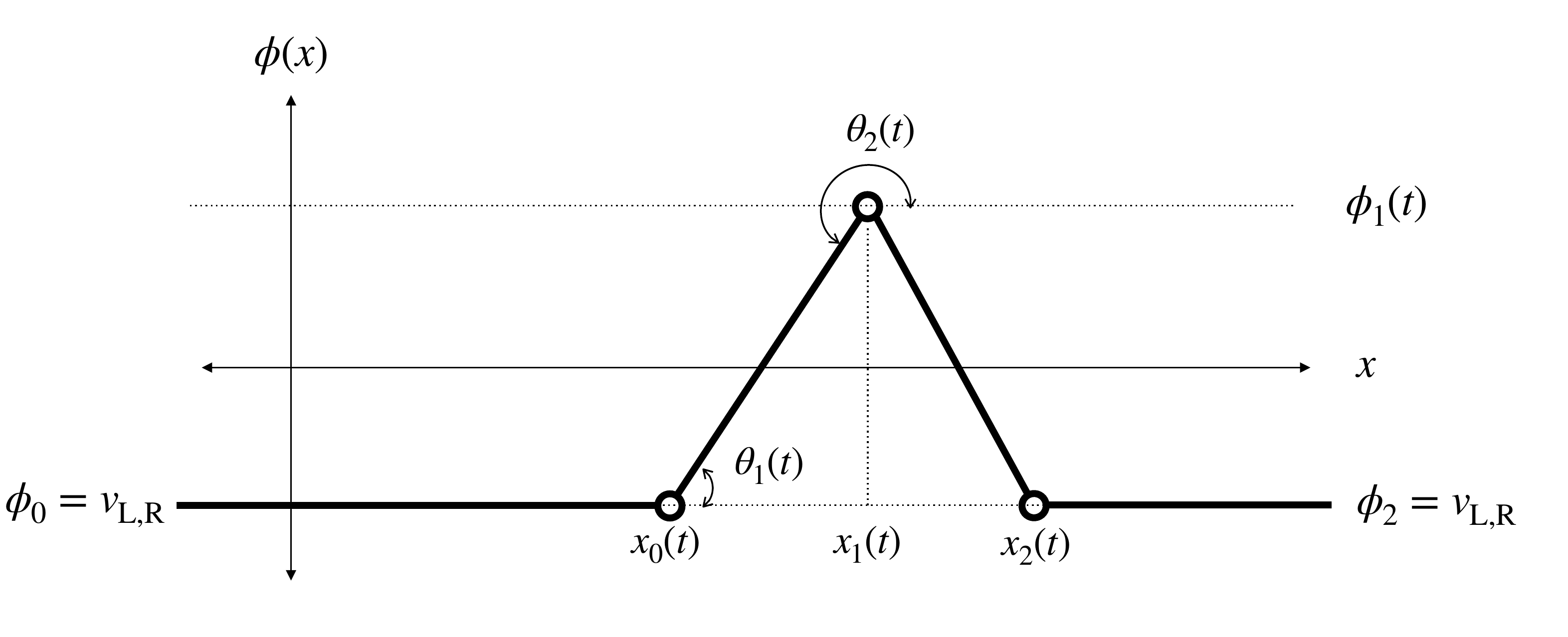}
\caption{\small Schematic depiction of $N=2$ mech-oscillon.}
\label{fig:1s4}
\end{center}
\end{figure}

In this section, we investigate non-topological configurations that have $v_{\rm L} = v_{\rm R} \equiv v$. We call them generically `mech-oscillons'. 
The simplest mech-oscillon is shown in Fig.~\ref{fig:1s4}.

\subsection{$N=2$: mech-field}

\begin{figure*}
\begin{center}
\includegraphics[width=0.95\textwidth]{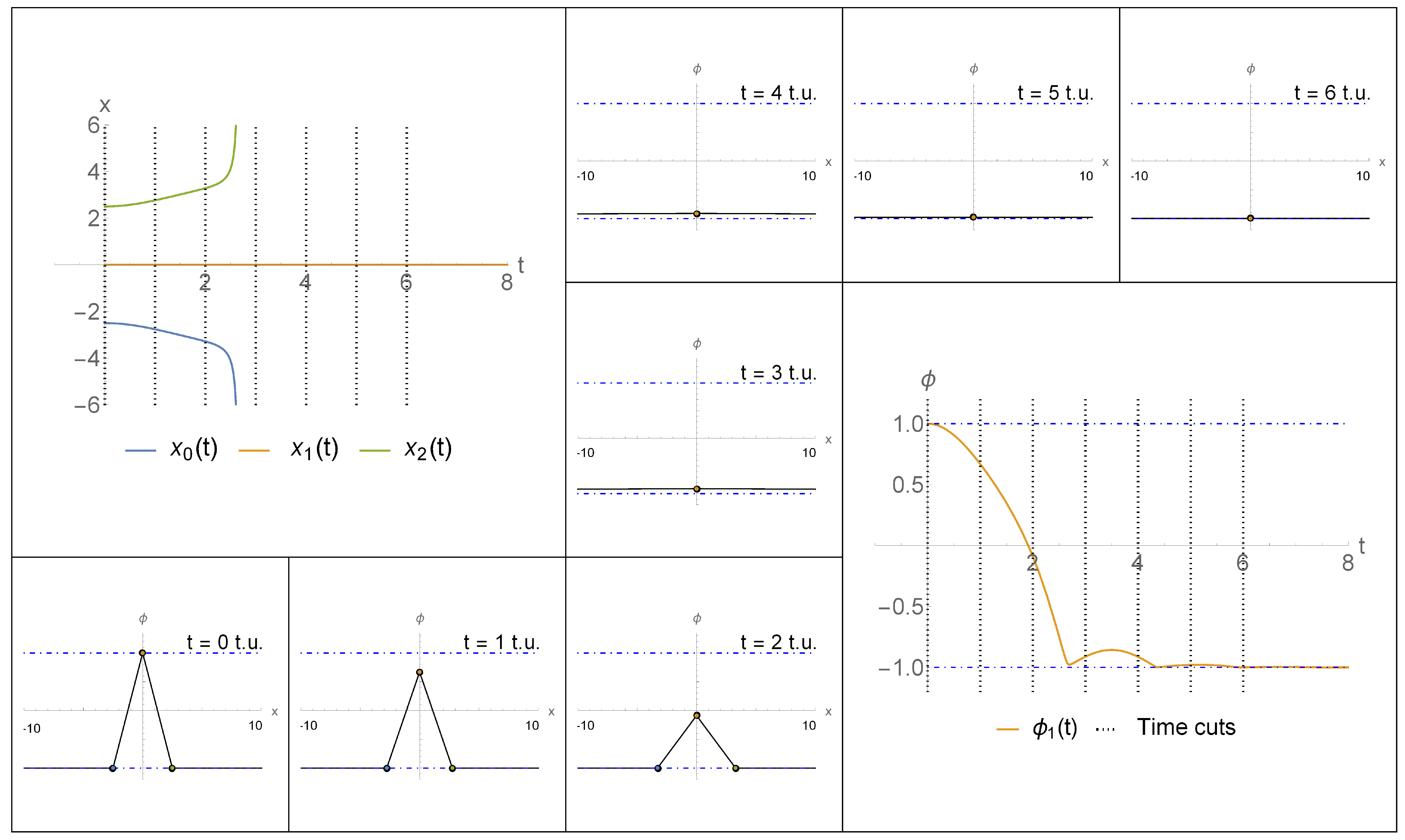}
\caption{\small An evolution of a mech-oscillon with initial conditions $R(0)=5$ and $A(0)=2$.}
\label{fig:2s4}
\end{center}
\end{figure*}

\begin{figure*}
\begin{center}
\includegraphics[width=0.95\textwidth]{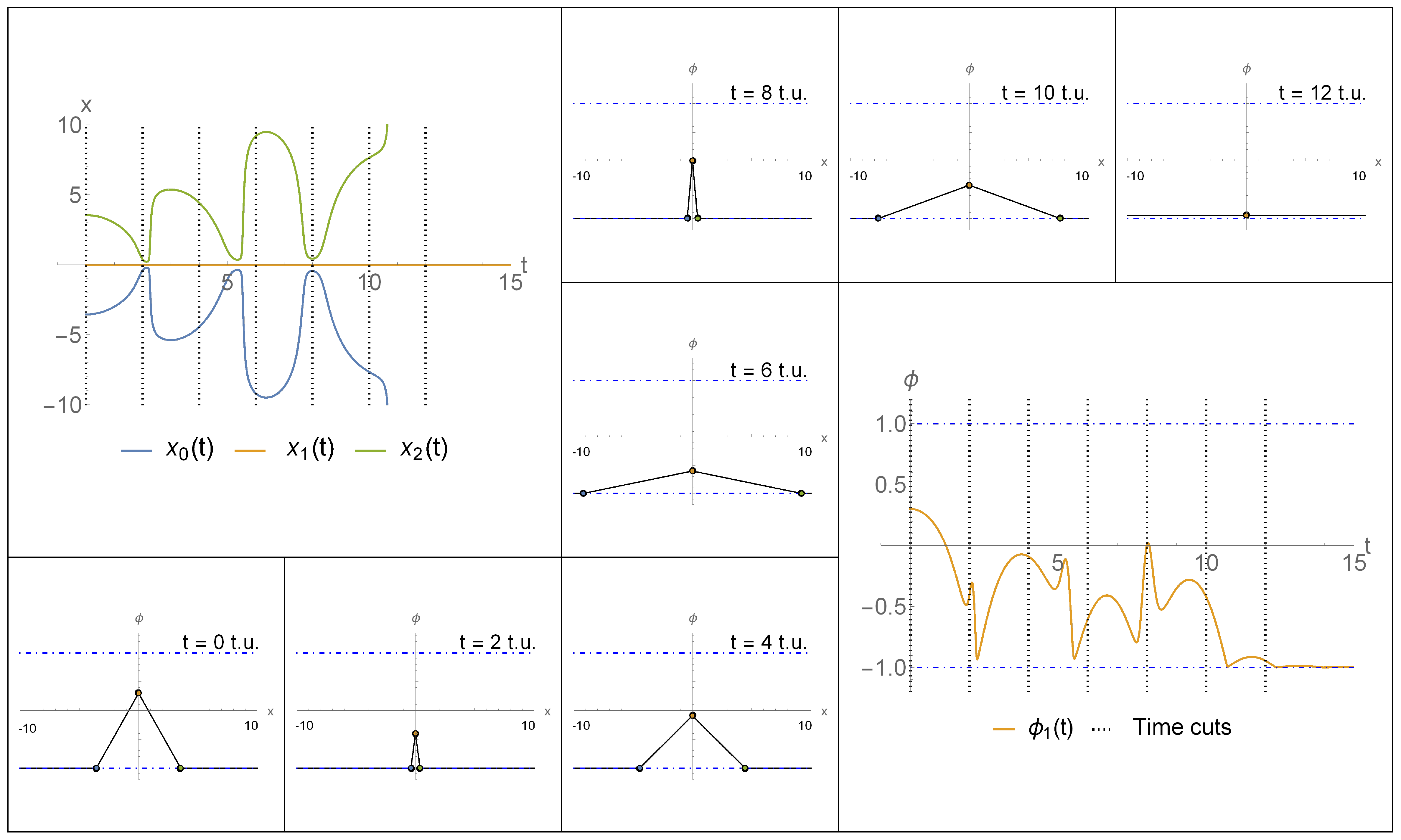}
\caption{\small An evolution of a mech-oscillon with initial conditions $R(0)=7.1$ and $A(0)=1.3$.}
\label{fig:4s4}
\end{center}
\end{figure*}

\begin{figure*}
\begin{center}
\includegraphics[width=0.95\textwidth]{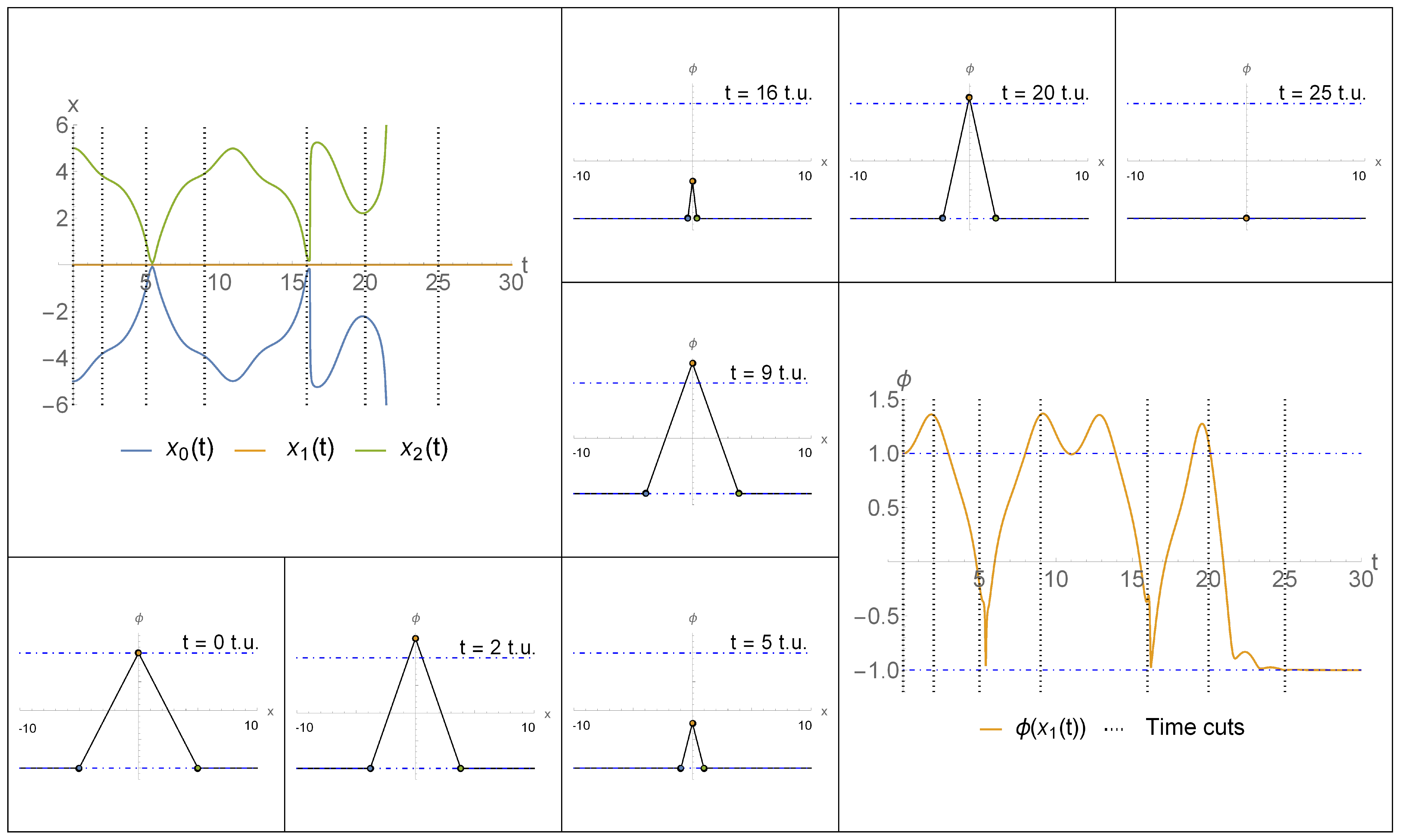}
\caption{\small An evolution of a mech-oscillon with initial conditions $R(0)=10$ and $A(0)=2$.}
\label{fig:3s4}
\end{center}
\end{figure*}

\begin{figure*}
\begin{center}
\includegraphics[width=0.95\textwidth]{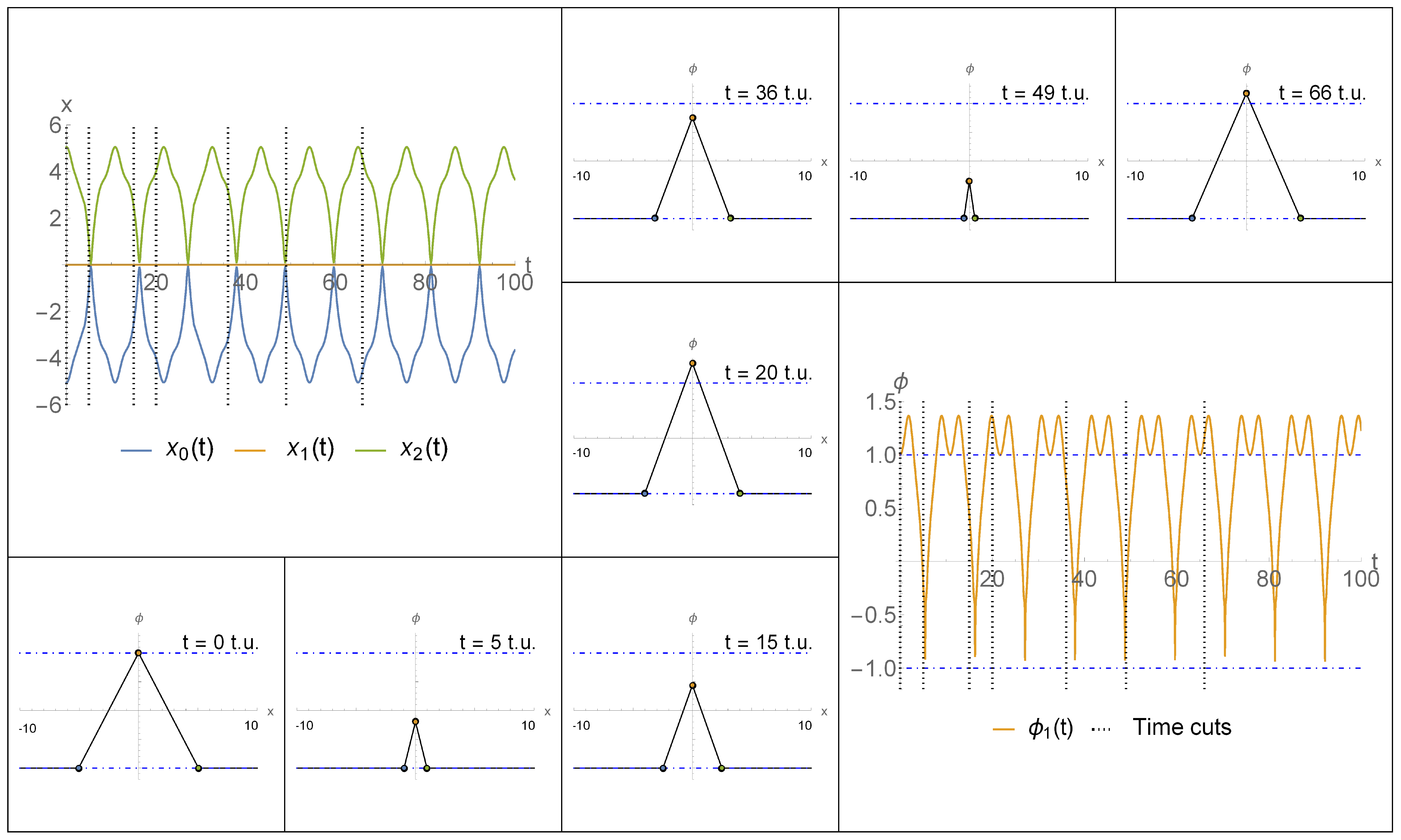}
\caption{\small An evolution of a mech-oscillon with initial conditions $R(0)=10.1$ and $A(0)=2$.}
\label{fig:5s4}
\end{center}
\end{figure*}

To make the analysis as simple as possible, let us investigate symmetric configuration, i.e.  a triangle of base length $R(t)$ placed on top of the vacuum $v$ with height $v+A(t)$ centered at the origin. In other words, we set
\begin{gather}
x_0(t) = -R(t)/2\,, \hspace{3mm}
x_1(t)= 0\,, \hspace{3mm}
x_2(t) = R(t)/2\,, \\
\phi_0(t) = \phi_2(t) = v\,, \hspace{3mm}
\phi_1(t) = v+A(t)\,.
\end{gather} 
This gives us
\begin{align}
L_M^{N=2} =& \frac{1}{6}R \dot A^2+\frac{1}{6}A\dot A\dot R+\frac{A^2}{6R}\bigl(\dot R^2-12\bigr)\nonumber \\
& -\frac{R}{A} \bigl({\mathcal V}(v+A)-{\mathcal V}(v)\bigr)\,.
\end{align}
In particular, for $\phi^4$ potential $V(\phi) = \frac{1}{2}\bigl(1-\phi^2\bigr)^2$we obtain (taking $v=-1$)
\begin{align}
3 L_M^{N=2} = & \frac{1}{2}R \dot A^2+\frac{1}{2}A\dot A\dot R+\frac{A^2}{2R}\dot R^2\nonumber \\ \label{eq:mechosclag}
& -\frac{6A^2}{R}-2 A^2 R+\frac{3}{2}A^3 R-\frac{3}{10}A^4R\,.
\end{align}
Note that $L_M^{N=2}$ has the same structure as in Eq.~\refer{eq:osclag}. Again, the similarity is due to the universality of CCMs for this type of background. Indeed, it is easy to see that the same terms with varying coefficients appear when using $\phi_{\rm bkg}= -1+A f\bigl(x/R\bigr)$, given some obvious convergence properties of $f$. Again, $N=2$ mech-kink falls into this class somewhat accidentally due to the imposed reflection symmetry. 
If we relax this restriction, i.e. $\Delta x_0 \not = \Delta x_1$, we no longer fit into this class, but a larger universality class of CCMs with three variables. However, we do not believe that this would yield qualitatively different dynamics.

We found it advantageous to use an exponential ansatz:\footnote{This ansatz removes the contact singularities and avoids $\Delta\phi_0 =0$ problem, making these coordinates especially convenient for numerical calculations. Notice that $A(t) \geq 0$ cannot go below zero. We numerically verified that even for general ansatz, the line $A(t)=0$ acts as a reflective barrier if approached either from above or below, hence our ansatz is not a loss of generality.}
\begin{equation}\label{eq:expo}
R(t) = \Exp{b(t)}\,, \hspace{5mm}
A(t) = \Exp{a(t)}\,.
\end{equation}
In these variables, the equations of motions in $\phi^4$ model reads:
\begin{align}
\ddot a  = & \dot b^2-\dot a^2+5 \Exp{a}-\frac{7}{5}\Exp{2a}-20 \Exp{-2b}-4\,, \\
\ddot b  = & -2\dot a\dot b-\dot b^2-\Exp{a}+\frac{2}{5}\Exp{2a}+16\Exp{-2b}\,, \\
 E_{\rm osc}  = & \frac{1}{6}\Exp{2a+b}\Bigl(\dot a^2+\dot a\dot b+\dot b^2 +4\Bigr)\nonumber \\ &+2\Exp{2a-b}+\frac{1}{10}\Exp{4a+b}-\frac{1}{2}\Exp{3a+b}\,.
\end{align}
where $E_{\rm osc} $ is the mech-oscillon's energy. The momentum is $P_{\rm osc} =0$ by construction.

\begin{figure}
\begin{center}
\includegraphics[width=\columnwidth]{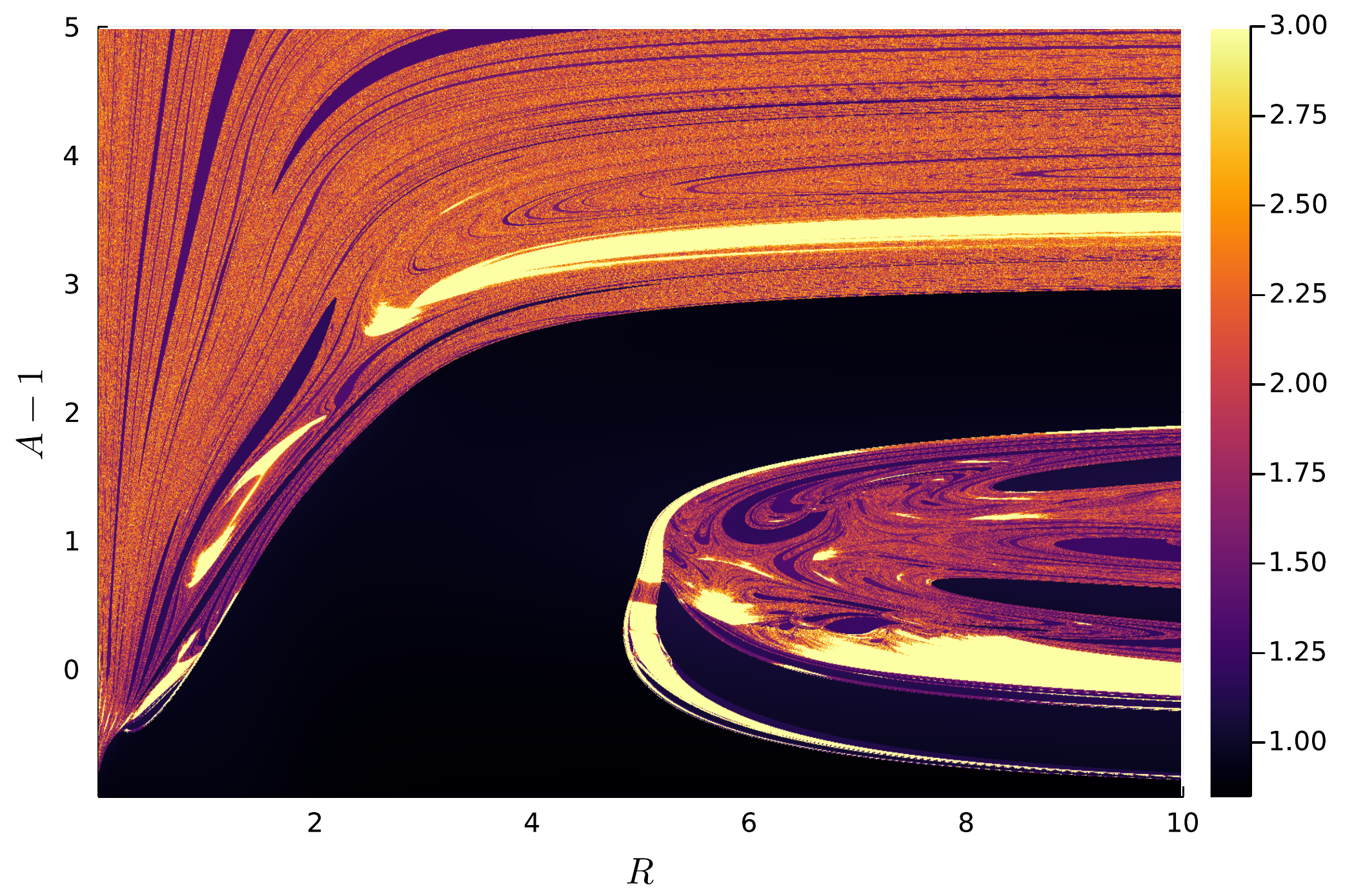}
\caption{\small A `map' of the lifetimes for $N=2$ mech-oscillons for initial values for $A$ and $R$. The colors represent $\log_{10}$ of the lifetime, while lifetimes beyond $10^3$ time units are all represented by the brightest color.}
\label{fig:map}
\end{center}
\end{figure}

We plot several solutions with varying initial conditions in Figs.~\ref{fig:2s4}-\ref{fig:5s4} that illustrate a typical behavior of a mech-oscillon. Namely, there is an initial quasi-periodic, chaotic phase followed by a decaying phase, in which the mech-oscillon rapidly collapses to vacuum. 
In other words, the mech-oscillon seems to have a well-defined lifetime, whose duration is very sensitive to initial conditions.  Indeed, the lifetime of a mech-oscillon can range from very short (Fig.~\ref{fig:2s4}) to extremely long (Fig.~\ref{fig:5s4}).\footnote{ Despite our best efforts, we did not observe the decay of this mech-oscillon. We only know that its lifetime must be longer than $\sim 74000$ time units.  In fact, there is a common understanding that in 1+1 dimensions, oscillons can have infinite lifetimes.} In Fig.~\ref{fig:map}, we display a `map' of  lifetime's dependence on the initial height and length of the triangle. Each color corresponds to a particular value of a $\log_{10}$ of the lifetime.

Lastly, we can understand the decaying phase by studying the asymptotic properties. It is easy to show that as $t\to \infty$ we have
\begin{equation}
a(t) \sim -\sqrt{\frac{1}{3}V^{\prime\prime}(v)}t\,, \hspace{2mm}
b(t) \sim 2\sqrt{\frac{1}{3}V^{\prime\prime}(v)}t\,, 
\end{equation}
In other words, the height of the mech-oscillon exponentially decays, while its width exponentially grows. Also notice that $b+2a$ remains constant. 

%\begin{figure}
%\begin{center}
%\includegraphics[width=\columnwidth]{Lifemap1.pdf}
%\caption{\small A detail of the `island' from the previous picture.}
%\label{fig:map1}
%\end{center}
%\end{figure}

\subsection{$N\geq 3$ mech-oscillons}

Mech-oscillons display a range of behaviors for $N\geq 3$. Among others, the most significant is mech-$K\bar K$ pair production. 

We can observe it already for $N=3$, where the mech-$K\bar K$ pair immediately fly apart and decouple from each other as the middle segment falls onto the vacuum. More interestingly, for $N\geq 4$, the created mech-$K\bar K$ pair remains connected via a mech-oscillon that facilitates the bouncing phenomenon (see Figs.~\ref{fig:6s4}-\ref{fig:8s4}). An interesting possibility suggests itself, namely a connection between the distribution of bouncing windows and the lifetime of mech-oscillon. As far as the authors are aware, this connection has not been explored in the field theory. 

It would be an interesting future project to map out the structure of the bouncing windows for $N=4$ mech-oscillons and compare it with Fig.~\ref{fig:map}. In that way, we may confirm this connection in our mech-models. The question of whether the same can be meaningfully established in field theory, however, is subtle. 

%For instance, here we study the evolution of mech-oscillons and not the scattering of mech-kinks, which may not be completely equivalent phenomena. Of course, there is the invariance under time-reversal that transforms solutions of decaying mech-oscillons into a scattering of excited mech-kinks. However, in our approach it is difficult to study the scattering of asymptotically free mech-kinks directly, due to the fact that they are  dynamically decoupled from each other.

%Another subtlety is that in field theory, during the $K\bar K$ collision the field becomes for an instant a pure vacuum. The same is not possible for a mech-field because the vacuum is dynamically decoupled.

In general, a nice feature of mech-models is the gradual discovery of new behaviors as $N$ increases. 
For instance, the sequential ejection of two  mech-$K\bar K$ pairs becomes possible only at $N=7$. 
For $N=8$, we can not only observe the ejection of two mech-$K\bar K$ pairs, but the trailing pair can also undergo a few bounces before escaping to infinity as Figs.~\ref{fig:9s4}-\ref{fig:10s4} illustrate.

\begin{figure*}
\begin{center}
\includegraphics[width=0.95\textwidth]{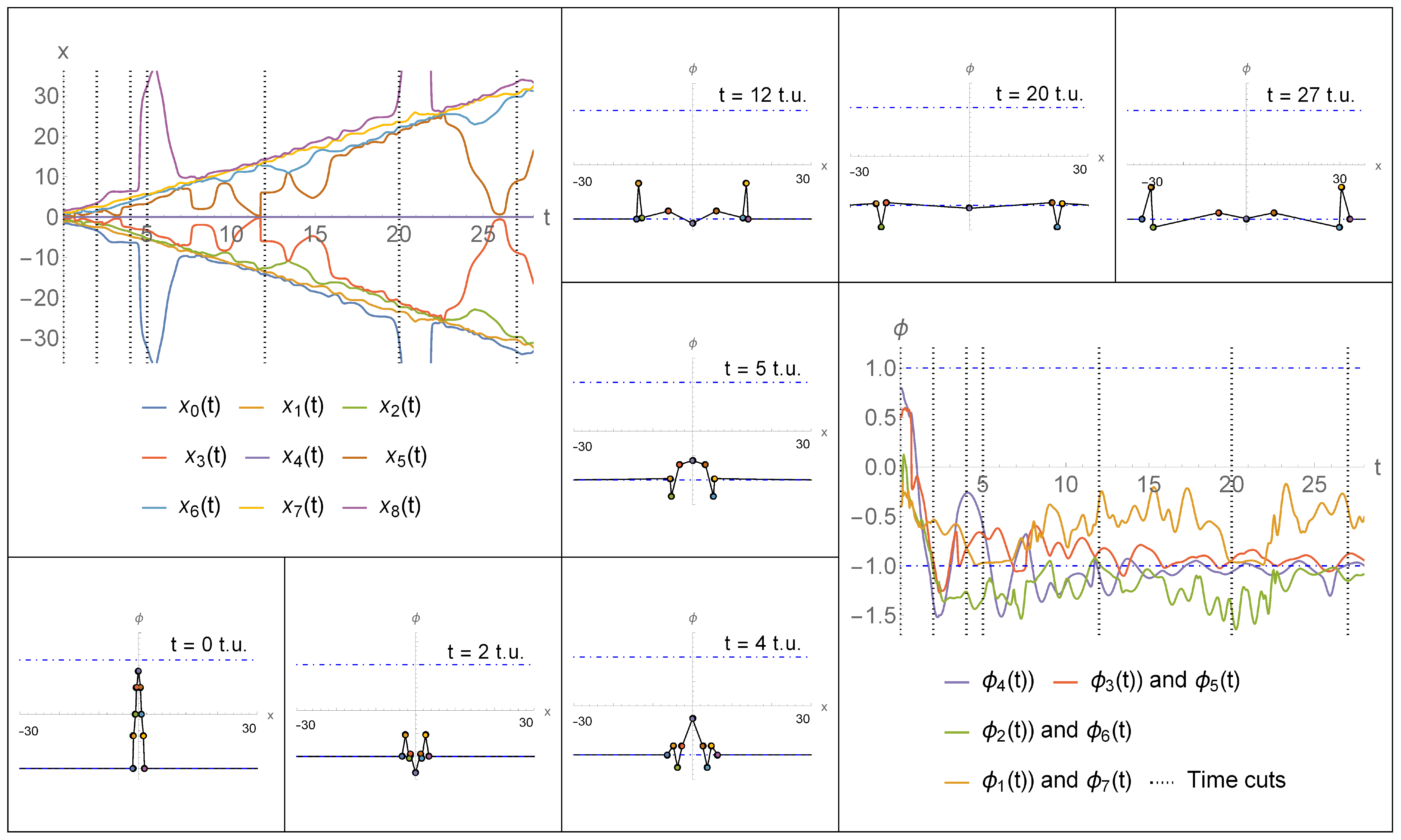}
\caption{\small Ejection of small-amplitude mech-oscillons for $N=8$ mech-field. Analog of radiative decay?}
\label{fig:13s4}
\end{center}
\end{figure*}

A different type of behaviour is the ejection of a pair of mech-oscillons, which we show on Figs.~\ref{fig:11s4}-\ref{fig:13s4}. In some instances (Figs.~\ref{fig:11s4}-\ref{fig:12s4}) the ejected mech-oscillons has large amplitudes that span the gap between the two vacua. These could be viewed as mech-bions, tight bound states of mech-$K\bar K$ pairs. In other instances (Fig.~\ref{fig:13s4}), the mech-oscillons have small amplitudes. More than anything else, they could represent mechanical analogs of a radiative decay.

For higher $N$ still (and sufficient energies) the mech-osillon may display a various combination of all these processes. In our analysis we investigated mech-fields up to $N=11$.

%%%%%%%%%%%%%%%%%%%%%%%%%%%%%%%%%%%%%%%%%%%%%%%%%%%%%%%%%%
%%%%%%%%%%%%%%%%%%%%%%%%%%%%%%%%%%%%%%%%%%%%%%%%%%%%%%%%%%
%%%%%%%%%%%%%%%%%%%%%%%%%%%%%%%%%%%%%%%%%%%%%%%%%%%%%%%%%%
%%%%%%%%%%%%%%%%%%%%%%%%%%%%%%%%%%%%%%%%%%%%%%%%%%%%%%%%%%
\section{Summary}
\label{sec:V}
%%%%%%%%%%%%%%%%%%%%%%%%%%%%%%%%%%%%%%%%%%%%%%%%%%%%%%%%%%
%%%%%%%%%%%%%%%%%%%%%%%%%%%%%%%%%%%%%%%%%%%%%%%%%%%%%%%%%%
%%%%%%%%%%%%%%%%%%%%%%%%%%%%%%%%%%%%%%%%%%%%%%%%%%%%%%%%%%
%%%%%%%%%%%%%%%%%%%%%%%%%%%%%%%%%%%%%%%%%%%%%%%%%%%%%%%%%%

In this paper we have presented a general-purpose $N$-point CCM for a simple scalar field theory in $1+1$ dimensions. It is based on the `mechanization' procedure in which a continuous field is replaced by a pice-wise linear function. The conceptual simplicity of our construction gave us algebraically tractable CCMs. Our numerical investigations indicate qualitative agreement between phenomena observed in mech-models and the field theory.

The question of quantitative agreement is left as a future task. At this point, we may only resort to a hand-waving statement that mech-field should resemble continuous theory more and more as $N$ increases. 

The most useful aspect of our approach is the natural ordering in the complexity of behaviors. As we have seen, the exploration of mech-models with an increasing number of joints gradually opens new dynamical modes of the mech-field.

Starting at $N=1$, we have `discovered' a mech-kink that behaves as a relativistic particle. Furthermore, the corresponding mech-model \refer{eq:bpsmechkinklag} turns out to be the same term-wise as relativistically covariant CCM based on position and scaling modulus \refer{eq:bpskinklag}. Let us point out that this result was given us \emph{for free} without any attempt to recover lost Lorentz covariance that motivates its construction in the field theory \cite{Manton:2020onl}. 

At $N=2$, we have found a mech-oscillon and its tendency to suddenly decay into a vacuum after a period of time that sensitively depends on initial conditions (see Fig.~\ref{fig:map}). The corresponding CCM \refer{eq:mechosclag} is, again, structurally the same as the field-theoretical one \refer{eq:osclag}.

%new
%Let us, however, point out that contrary to mech-kink that mimics its field-theoretical prodigy quite well, the mech-oscillon qualitatively differs from field-theoretical oscillons in important respects. In particular, the field value of $N=2$ mech-oscillon cannot cross the vacuum on which it is based (in our notation for double-well model it is the value $A=0$). This is because the completely flat mech-oscillon is an \emph{exact} solution of the equations of motion, namely the vacuum itself. Since the solutions cannot cross, the vacuum represents a separatrix between $A>0$ and $A<0$ solutions. Other difference is the fact that mech-oscillons can have very long lifetimes even for large amplitudes, while this is not so in field theory. The explanation for this discrepancy is clearly the absence of radiation modes for $N=2$ mech-oscillons, that are crucial for the life-time of field-theoretical oscillons. From these observations it is clear that $N=2$ mech-oscillon is a very different entity than oscillons in field theory. However, both issues become less severe for more elaborate mech-oscillons. With multiple additional segments, mech-oscillons can cross the vacuum (as long as not all segments are flat) and there exists radiation modes. We plan to investigate the life-time of mech-oscillons more closely in the future. 

For $N=2$ mech-kinks, we observed a phenomenon of joint-ejection (Fig.~\ref{fig:eight}) that seems to be endemic for all configurations with $N\geq 2$ (Fig.~\ref{fig:nine}). The joint-ejection exemplifies a general tendency that is characteristic across our numerical data. Namely, the proclivity of initially tightly bound mech-field to disintegrate over time into most basic configurations, such as $N=1$ mech-kinks and $N=2$ mech-oscillons (or their pairs), that separate and gradually decouple from each other.

\begin{figure*}
\begin{center}
\includegraphics[width=0.95\textwidth]{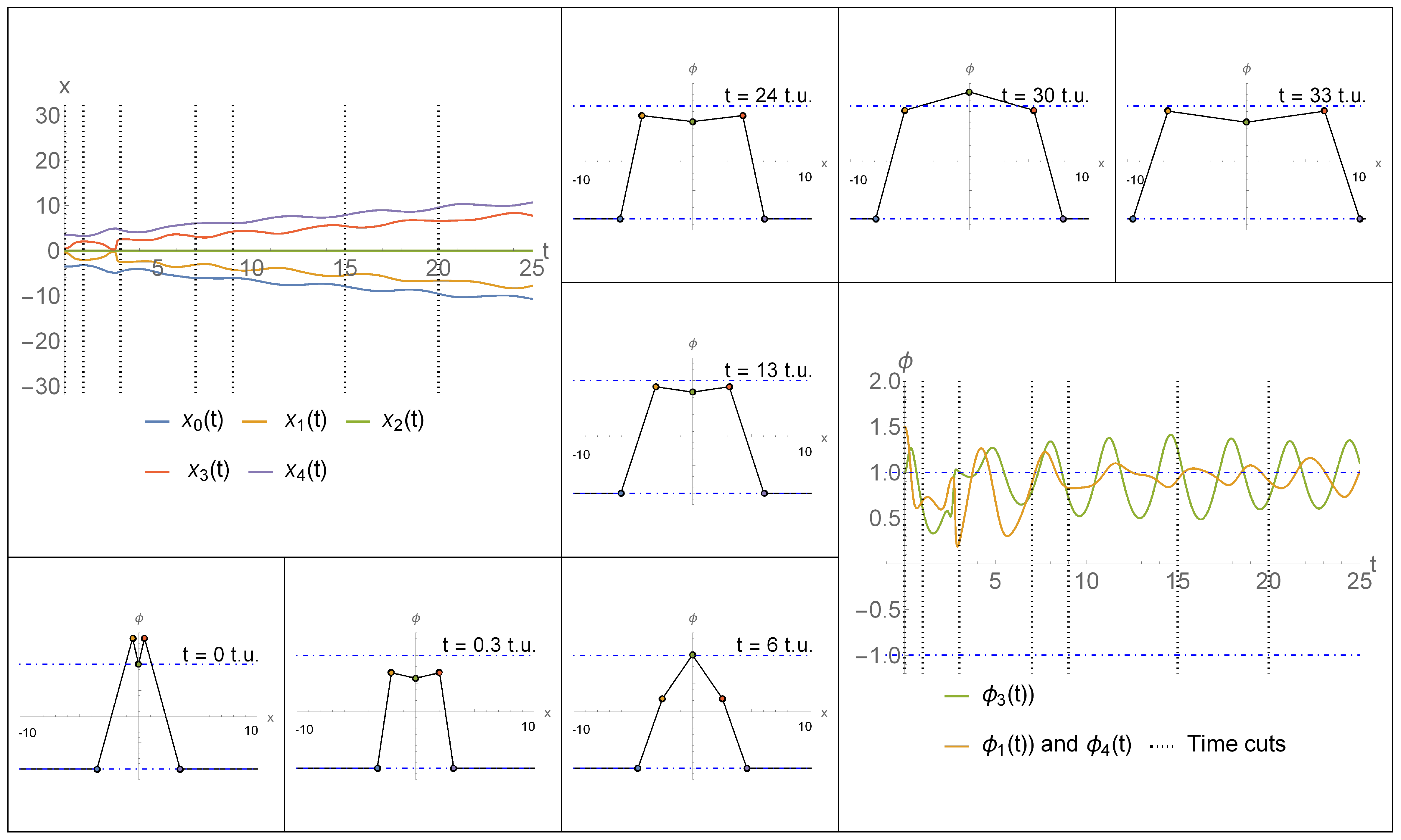}
\caption{\small A mech-$K\bar K$ ejection from initial $N=4$ mech-oscillon leaving behind $N=2$ mech-oscillon.}
\label{fig:6s4}
\end{center}
\end{figure*}

\begin{figure*}
\begin{center}
\includegraphics[width=0.95\textwidth]{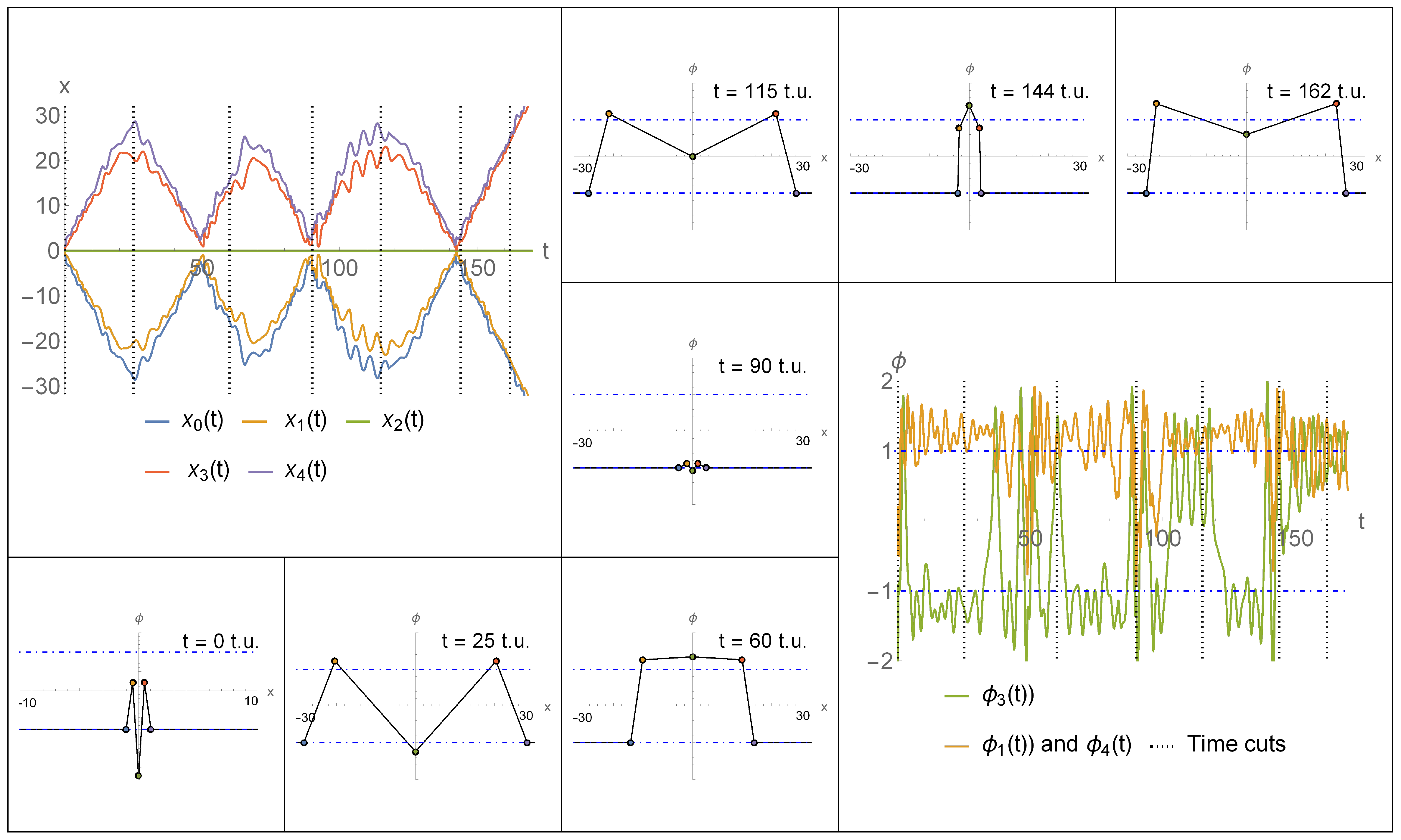}
\caption{\small Initial `bounces' before a mech-$K\bar K$ ejection for $N=4$ mech-oscillon .}
\label{fig:7s4}
\end{center}
\end{figure*}

\begin{figure*}
\begin{center}
\includegraphics[width=0.95\textwidth]{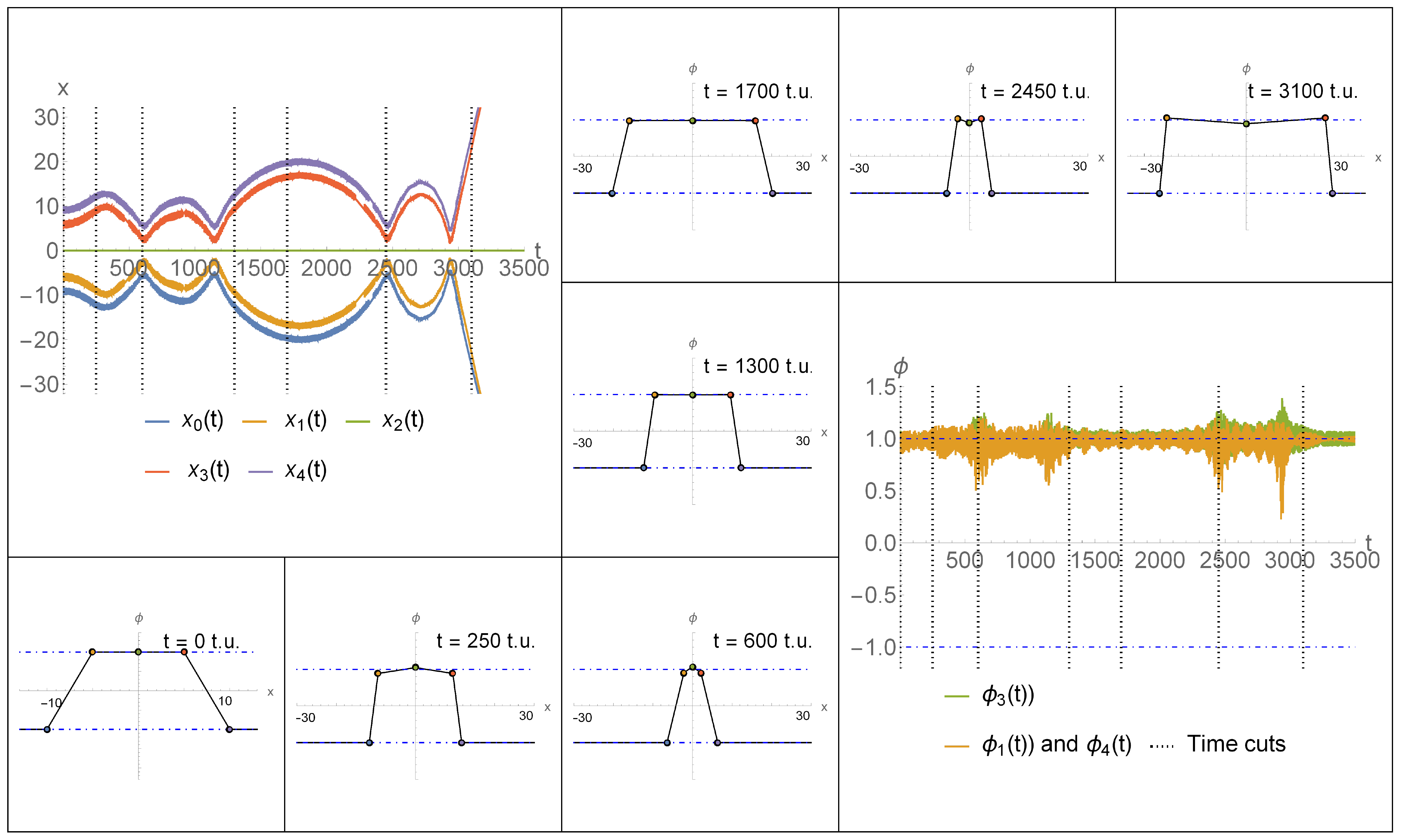}
\caption{\small Sequence of long bounces before a mech-$K\bar K$ ejection for $N=4$ mech-oscillon.}
\label{fig:8s4}
\end{center}
\end{figure*}

The production of mech-$K\bar K$ pairs can be seen already at $N=3$ level, but it is for $N=4$ mech-oscillons that the phenomenon of bouncing starts to manifest (Figs.~\ref{fig:6s4}, \ref{fig:7s4} and \ref{fig:8s4}). As the mech-$K\bar K$ pair is bound together via a mech-oscillon, a curious connection between phenomena of bouncing and mech-oscillon's lifetime suggests itself. We have not investigated this possibility in detail, however, and it remains an interesting future work. 

\begin{figure*}
\begin{center}
\includegraphics[width=0.95\textwidth]{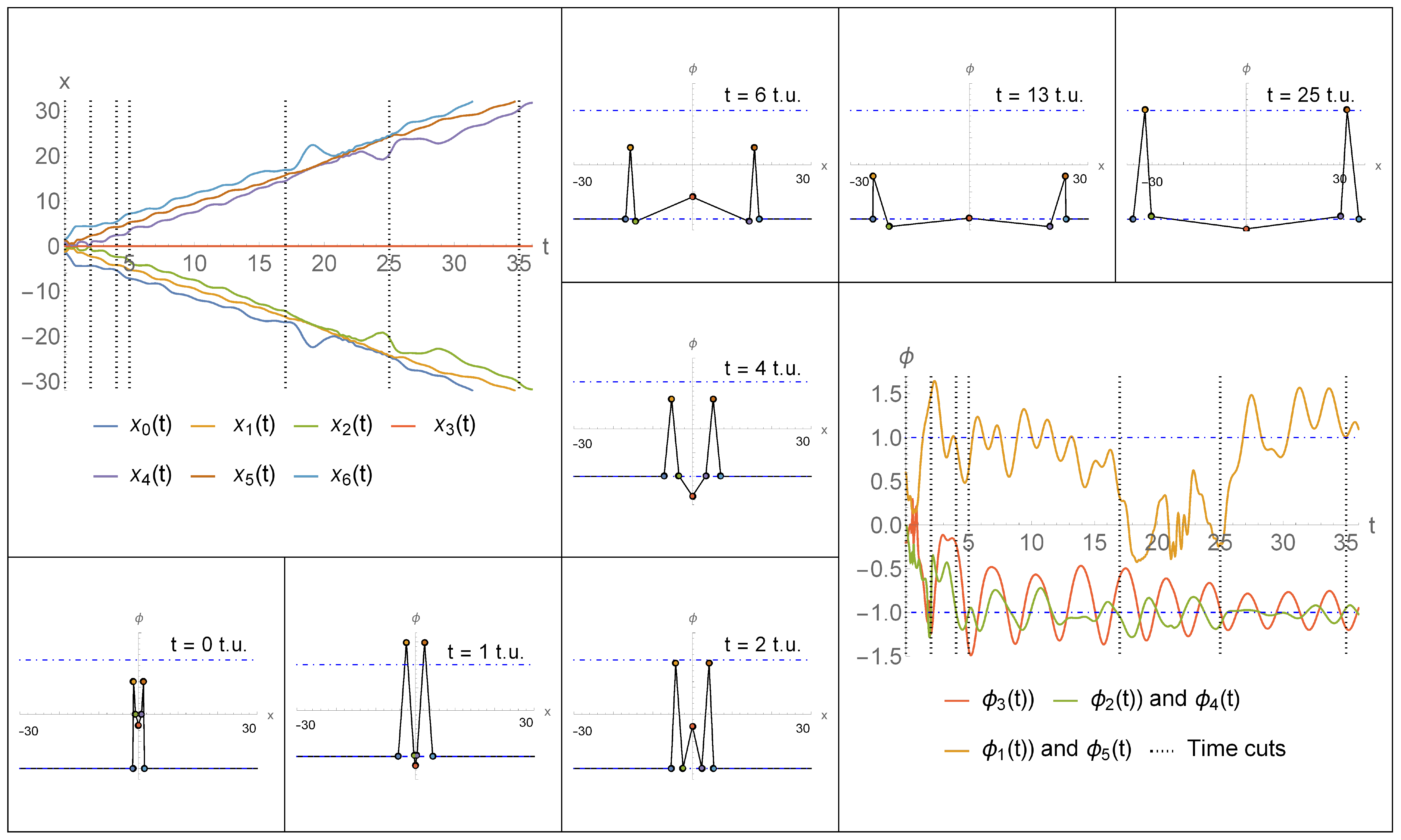}
\caption{\small Ejection of a pair of mech-oscillons for $N=6$ mech-field.}
\label{fig:11s4}
\end{center}
\end{figure*}

\begin{figure*}
\begin{center}
\includegraphics[width=0.95\textwidth]{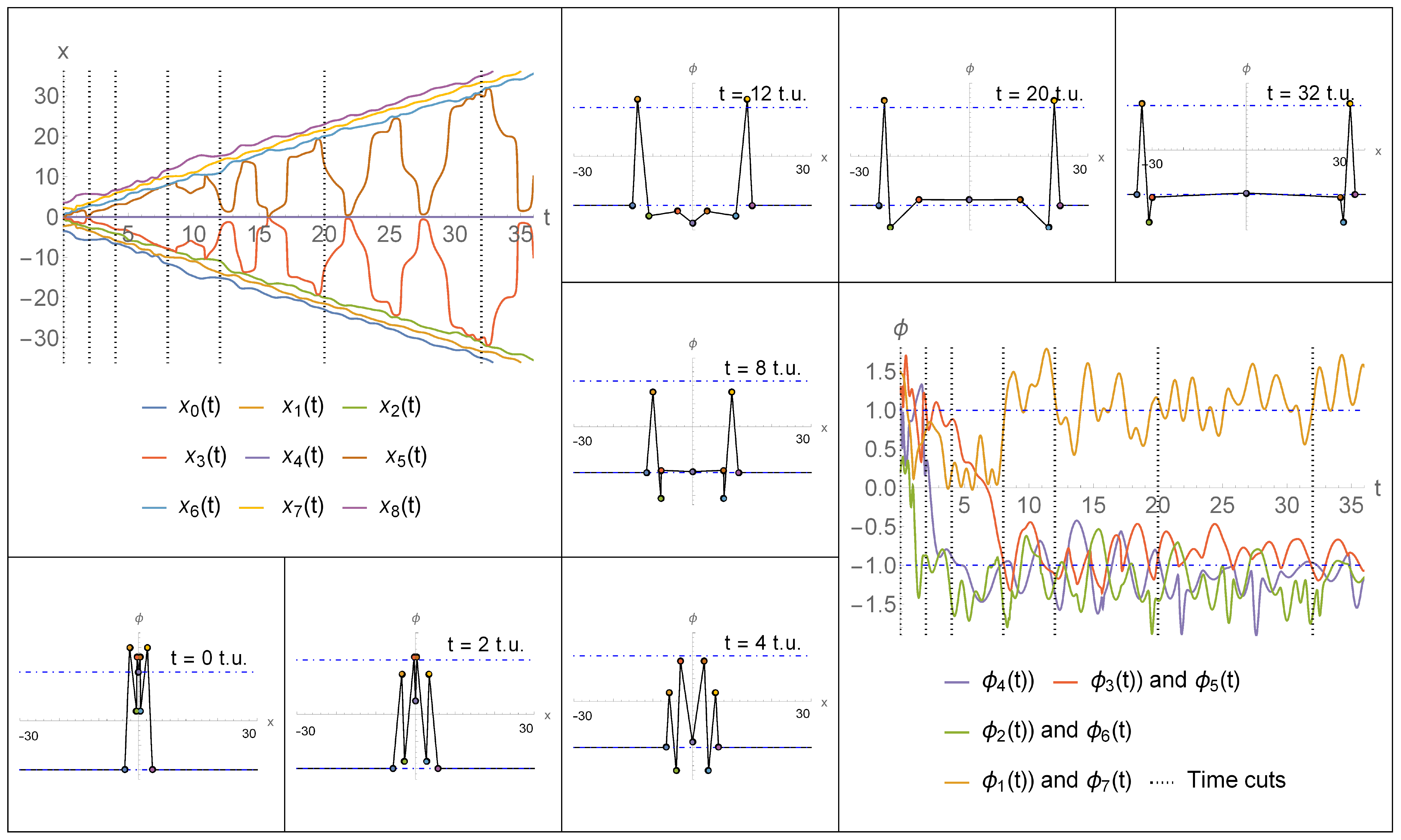}
\caption{\small Ejection of pair of large-amplitude mech-oscillons in $N=8$ configuration, illustrating production of mech-bions (tight mech-$K\bar K$ bound states).}
\label{fig:12s4}
\end{center}
\end{figure*}

At $N=6$, yet another mode opens up, namely an ejection of a pair of mech-oscillons (Fig.~\ref{fig:11s4}). For $N=8$, we observed that the peaks of ejected mech-oscillons sometimes reach all the way to the second vacuum (Fig.~\ref{fig:12s4}). In such situations, it seems reasonable to view the ejecta as mech-bions -- tight mech-$K\bar K$ bound states. Other times, however, the mech-oscillons have very small amplitudes. These situations are perhaps analogs of field-theoretical radiation decay.   

\begin{figure*}
\begin{center}
\includegraphics[width=0.95\textwidth]{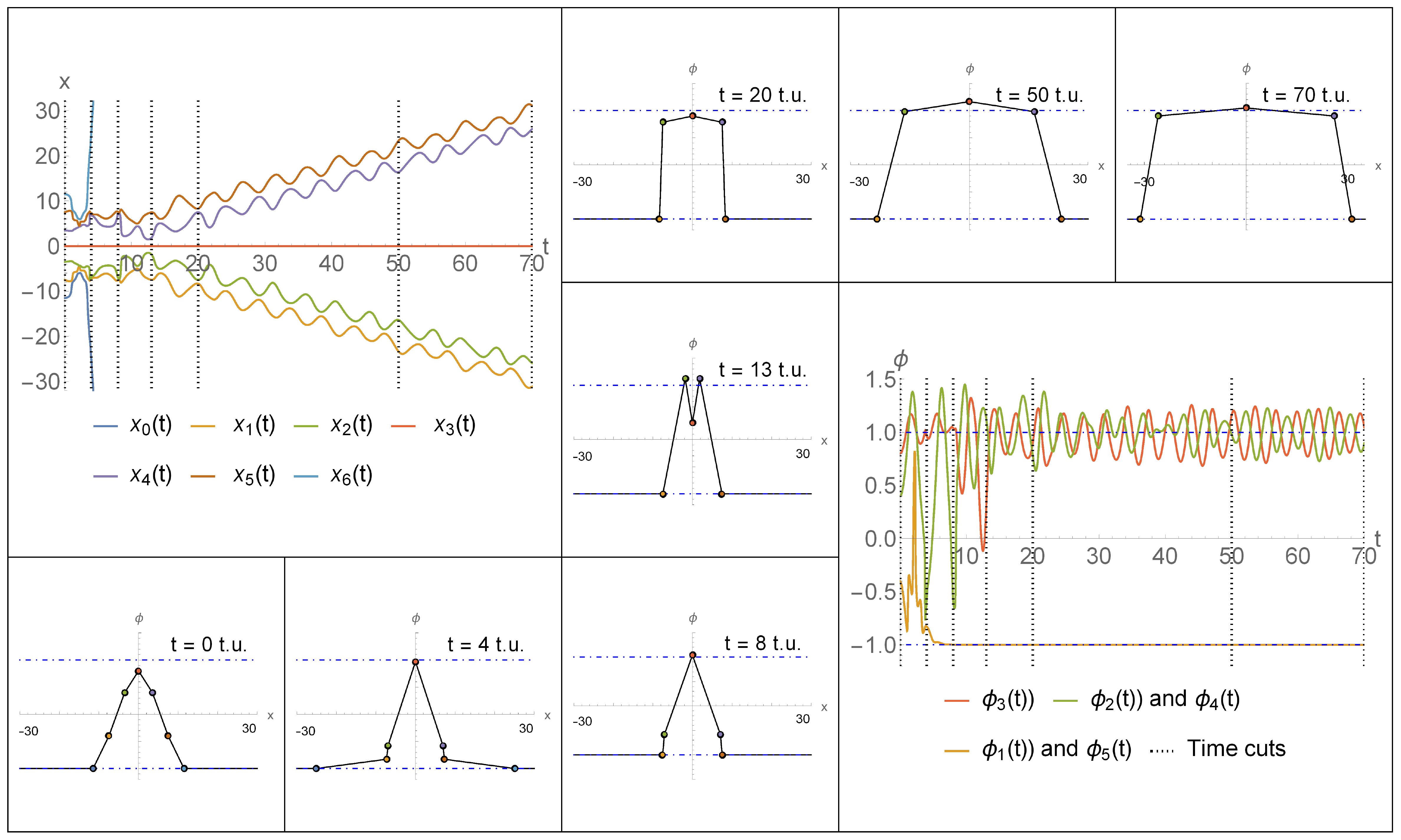}
\caption{\small An initial ejection of outermost joints followed by ejection of a highly-excited mech-$K\bar K$ pair for $N=6$ mech-oscillon.}
\label{fig:6p5s4}
\end{center}
\end{figure*}

\begin{figure*}
\begin{center}
\includegraphics[width=0.95\textwidth]{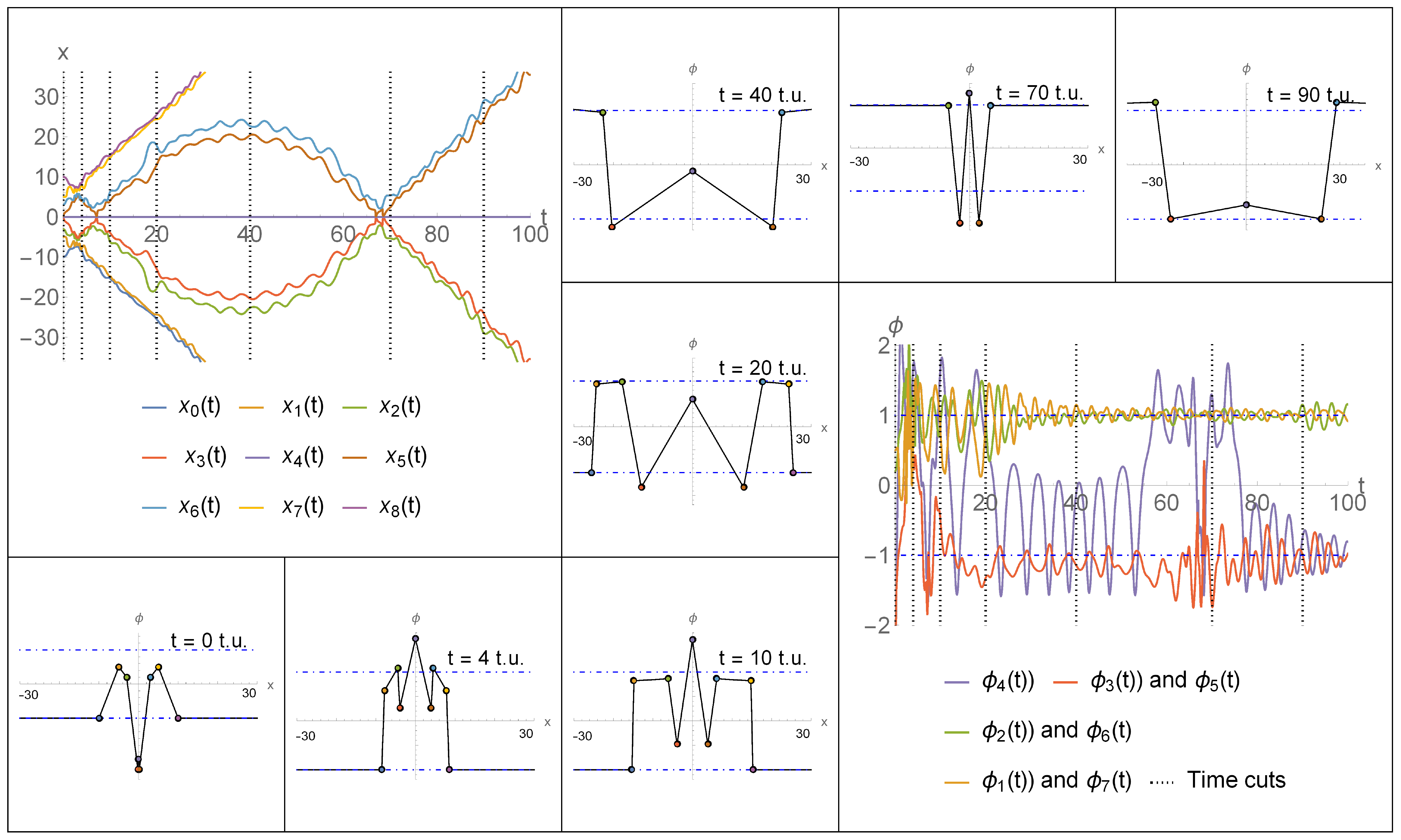}
\caption{\small Ejection of two mech-$K\bar K$ pairs with two intermediary bounces of the second pair for $N=8$ mech-oscillon.}
\label{fig:9s4}
\end{center}
\end{figure*}

\begin{figure*}
\begin{center}
\includegraphics[width=0.95\textwidth]{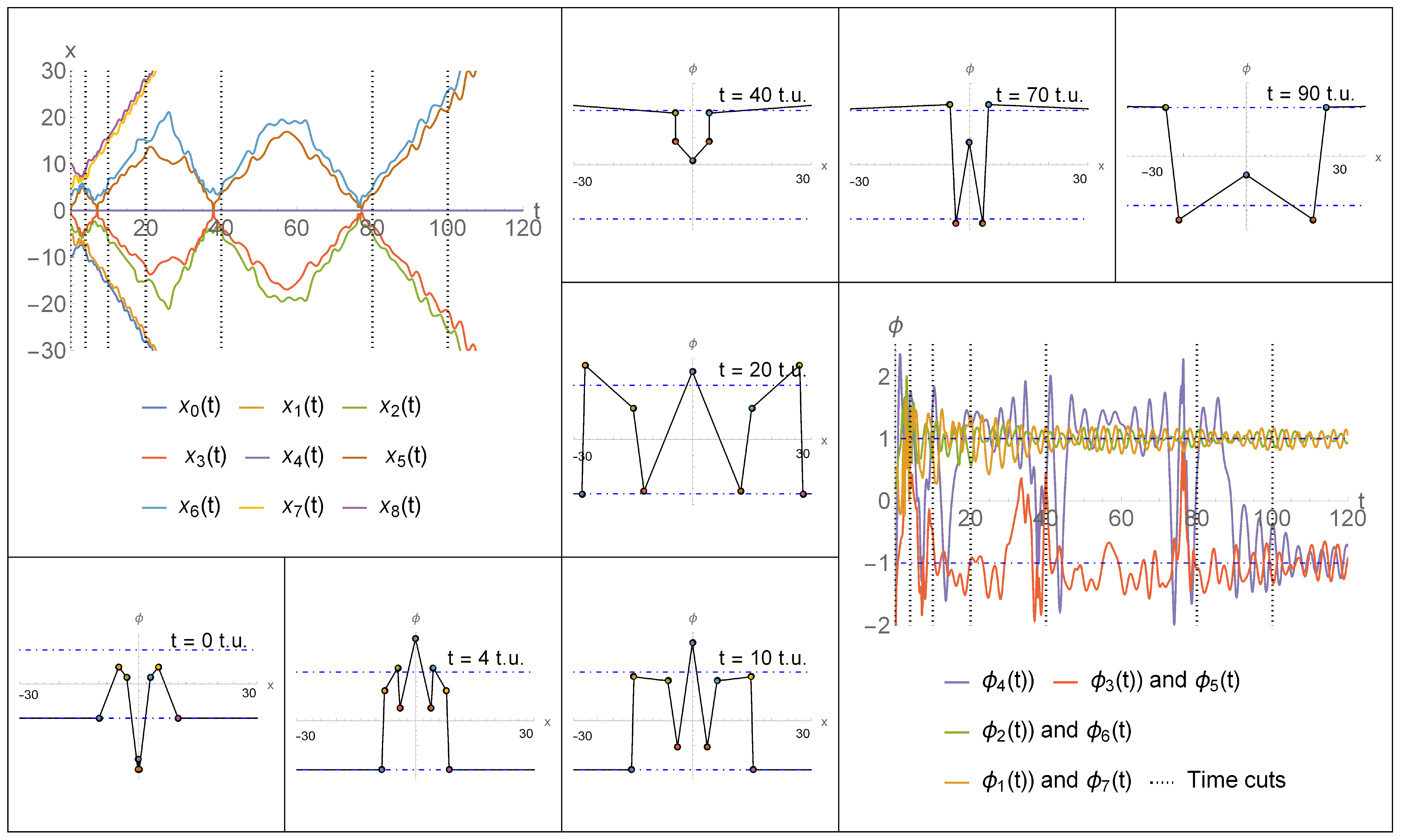}
\caption{\small Ejection of two mech-$K\bar K$ pairs with three intermediary bounces of the second pair for $N=8$ mech-oscillon.}
\label{fig:10s4}
\end{center}
\end{figure*}

Of course, arbitrary combinations of the above processes can occur in sequence for sufficiently high $N$ (Figs.~\ref{fig:6p5s4}, \ref{fig:9s4} and \ref{fig:10s4}). 

Let us stress that mechanization should be viewed as a proof-of-concept rather than a serious attempt for general-purpose CCMs. Indeed, there are issues with our construction. The most glaring one is the geodetic incompleteness of the moduli space, which is perhaps the largest source of quantitative disagreement. 

As we discussed, singularities arise whenever the distance between two joints becomes zero, i.e. $\Delta x_a=0$, or a middle segment becomes flat, i.e. $\Delta \phi_a =0$. The second type of singularities introduces an unexpected practical problem: it is quite challenging to investigate mech-$K\bar K$ scattering directly. This is because the configuration of initially separated mech-kink and anti-mech-kink with a flat segment in between is dynamically decoupled. 
Indeed, such a configuration is an exact solution of the equations of motion describing free particles. However, upon contact, we arrive at $\Delta x_a =0$ type singularity, and the equations of motion break down.
For this reason, we have mostly investigated the evolution of large $N$ mech-oscillons that provide an indirect way of studying  the scattering of mech-kinks.

The resolution of moduli space singularities would be a fundamental step forward. At present, it is not clear to us how we should accomplish it. It is telling, however, that the singularities appear whenever the mech-field suddenly changes its (effective) number of joints. For example, one may continue beyond a singular collision of a free-moving mech-$K\bar K$ pair, which is a $N=3$ configuration, by replacing it with a $N=2$ mech-oscillon. Although intuitive, it is difficult to realize this approach in practice.

In other words, we should figure out how to dynamically connect different $N$-sectors. Making the number of particles vary would also provide a step towards restoring explicit Lorentz covariance. There is an obvious way to achieve this that we already know, i.e. taking $N\to \infty$ and reintroducing back continuous field. Whether there exists a middle ground where $N$ would remain finite and discrete dynamical variable remains a tantalizing possibility.

\acknowledgments

O. N. K. would like to thank Luk\'a\v{s} Rafaj for useful assistance.
The authors are also indebted to A.~Wereszczynski, T.~Romanczukiewicz and K.~Oles for discussions and useful feedback. Also we would like to thank T.~Romanczukiewicz for his help with our numerical code. 
F. B. would like to express his acknowledgment for the institutional support
of the Research Centre for Theoretical Physics and Astrophysics, Institute of Physics, Silesian University
in Opava and to the Institute of Experimental and Applied Physics, Czech Technical University in Prague.
This work was supported by the Student Grant Foundation of the Silesian University in Opava, Grant No. SGF/3/2021, which was realized within the EU OPSRE project entitled "Improving the quality of the internal grand scheme of the Silesian University in Opava", reg. number: CZ.02.2.69/0.0/0.0/19\_073/0016951.

%%%%%%%%%%%%%%%%

%\newpage

%\appendix

%\section*{Gallery}

%\begin{figure*}
%\begin{center}
%\includegraphics[width=0.95\textwidth]{fig08.pdf}
%\caption{\small An example of joint-ejection for $N=2$ mech-kink, where the third joint escapes to $\infty$.}
%\label{fig:eight}
%\end{center}
%\end{figure*}

%\begin{figure*}
%\begin{center}
%\includegraphics[width=0.95\textwidth]{fig09.pdf}
%\caption{\small A symmetric example where the outermost joints of $N=4$ mech-kink get ejected, leaving behind slightly excited $N=2$ mech-kink.}
%\label{fig:nine}
%\end{center}
%\end{figure*}

%\begin{figure*}
%\begin{center}
%\includegraphics[width=0.95\textwidth]{2s4.pdf}
%\caption{\small An evolution of a mech-oscillon wit initial conditions $R(0)=5$ and $A(0)=2$.}
%\label{fig:2s4}
%\end{center}
%\end{figure*}

%\begin{figure*}
%\begin{center}
%\includegraphics[width=0.95\textwidth]{3s4.pdf}
%\caption{\small An evolution of a mech-oscillon wit initial conditions $R(0)=10$ and $A(0)=2$.}
%\label{fig:3s4}
%\end{center}
%\end{figure*}

%\begin{figure*}
%\begin{center}
%\includegraphics[width=0.95\textwidth]{4s4.pdf}
%\caption{\small An evolution of a mech-oscillon wit initial conditions $R(0)=7.1$ and $A(0)=1.3$.}
%\label{fig:4s4}
%\end{center}
%\end{figure*}

%\begin{figure*}
%\begin{center}
%\includegraphics[width=0.95\textwidth]{5s4.pdf}
%\caption{\small An evolution of a mech-oscillon with initial conditions $R(0)=10.1$ and $A(0)=2$.}
%\label{fig:5s4}
%\end{center}
%\end{figure*}

\end{document}